\documentclass[prx,aps,amsmath,amssymb,twocolumn,showpacs,superscriptaddress,10pt]{revtex4-2}
\usepackage{graphicx}
\usepackage{bm}
\usepackage{bbold}

\usepackage{nicematrix}
\usepackage[colorlinks,bookmarks=true,citecolor=blue,linkcolor=blue,urlcolor=blue]{hyperref}
\usepackage{siunitx}

\usepackage[dvipsnames]{xcolor}

\newcommand{\tr}{\mathrm{tr}}
\newcommand{\upp}{\mathrm{p}}
\newcommand{\ups}{\mathrm{s}}
\newcommand{\ii}{\mathrm{i}}

\makeatletter
\renewcommand*{\fnum@figure}{{\normalfont\bfseries \figurename~\thefigure}}
\makeatother

\begin{document}
\title{Higher-order exceptional points in a multimode continuum optoacoustic system}

\author{Anton Montag}
\email{anton.montag@mpl.mpg.de}
\affiliation{Max Planck Institute for the Science of Light, 91058 Erlangen, Germany}
\affiliation{Department of Physics, Friedrich-Alexander-Universit\"at Erlangen-N\"urnberg, 91058 Erlangen, Germany}
\author{Julius T. Gohsrich}
\affiliation{Max Planck Institute for the Science of Light, 91058 Erlangen, Germany}
\affiliation{Department of Physics, Friedrich-Alexander-Universit\"at Erlangen-N\"urnberg, 91058 Erlangen, Germany}
\author{Quentin Levoy}
\affiliation{Max Planck Institute for the Science of Light, 91058 Erlangen, Germany}
\affiliation{Laboratoire Collisions Agrégats Réactivité, Université de Toulouse, CNRS, 31062 Toulouse, France}
\author{Birgit Stiller}
\email{birgit.stiller@mpl.mpg.de}
\affiliation{Max Planck Institute for the Science of Light, 91058 Erlangen, Germany}
\affiliation{Institute of Photonics, Leibniz University Hannover, 30167 Hannover, Germany}
\affiliation{Department of Physics, Friedrich-Alexander-Universit\"at Erlangen-N\"urnberg, 91058 Erlangen, Germany}
\author{Flore K. Kunst}
\email{flore.kunst@mpl.mpg.de}
\affiliation{Max Planck Institute for the Science of Light, 91058 Erlangen, Germany}
\affiliation{Department of Physics, Friedrich-Alexander-Universit\"at Erlangen-N\"urnberg, 91058 Erlangen, Germany}
\date{\today}

\begin{abstract}
    Exceptional points appear in non-Hermitian systems as degeneracies, where not only eigenvalues but also eigenvectors coalesce. They are of great theoretical and experimental interest due to their exotic topological properties and enhanced sensitivity to perturbations. Experimental realizations of higher-order exceptional points, where more than two eigenvectors coalesce, rely on highly fine-tuned setups. Recently, stimulated Brillouin scattering has been employed to generate second-order exceptional points in a fabrication-free setup by leveraging off-resonant scattering. In this work we generalize this approach, and we develop an off-resonant, multimode theory for stimulated Brillouin scattering as an avenue towards realizing symmetry-induced exceptional points of any order. We present the experimental implementation of our program in an accompanying paper. Our multimode theory could also be employed in applications in optoacoustic sensing, synthetic neuromorphic computing, microwave photonic filters, and optoacoustic quantum signal processing.
\end{abstract}

\maketitle

\section{Introduction}

Non-Hermitian approaches have proven themselves to be extremely useful in the field of optics~\cite{El-Ganainy2018,ozdemir_paritytime_2019,Miri2019}, where dissipation is ubiquitous, and have seen a wide span of attention from various fields in physics over the last decade~\cite{Bergholtz2021, Ashida2020}. 
Of particular interest is the appearance of exceptional points~\cite{KatoBook, Heiss2012} in the spectrum of non-Hermitian matrices~\cite{Miri2019}, which are points at which not only the eigenvalues but also the eigenvectors coalesce. 
Exceptional points of order two (EP2s), at which two eigenvalues and two eigenvectors coalesce, are abundant as one generally only needs to satisfy two real constraints~\cite{Berry2004, Carlstrom2018}.
EP2s have been observed in various experiments ranging from optical microring cavities~\cite{Brandstetter2014} and multilayer passive structures~\cite{Feng2013} to optical waveguides~\cite{Guo2009,Ruter2010,Feng2013a} and photonic crystals~\cite{Zhen2015}. 
These experiments reveal exotic phenomena associated with these EP2s such as unidirectional invisibility~\cite{Lin2011}, unidirectional lasing~\cite{Peng2016}, and tuning of the group velocity of light pulses~\cite{Goldzak2018}.

Exceptional points of higher order may also appear but require the tuning of more parameters. 
In general, an exceptional point of order $N$ (EP$N$) appears upon satisfying $2(N-1)$ real constraints, such that one typically has to resort to fine-tuning to find them~\cite{Delplace2021,Sayyad2022}. 
However, recent works show that symmetries, or more generally similarities~\cite{Montag2024,Montag2024_2,Montag2025}, reduce the number of constraints to realize exceptional points.
Higher-order exceptional points with \mbox{$N\geq3$} have been observed in coupled optical microcavities \cite{hodaei_enhanced_2017,jahangiri_observation_2025}, coupled acoustic cavities \cite{tang_exceptional_2020,tang_realization_2023}, an optomechanical cavity~\cite{Patil2022}, radio-frequency electric circuits \cite{yin_high-order_2023}, single-photon interferometry \cite{wang_experimental_2023}, nitrogen-vacancy centers \cite{wu_third-order_2024}, Bose-Einstein condensates \cite{wang_exceptional_2024}, superconducting circuits \cite{han_measuring_2024,zhang_topological_2025,zhang_experimental_2025}, and trapped ions \cite{chen_quantum_2025,li_programmable_2024}.
All the aforementioned experimental setups either use fabricated fine-tuned structures, or rely on a fixed discrete level structure with meticulously engineered couplings.
\begin{figure}
    \centering
    \includegraphics[width=\linewidth]{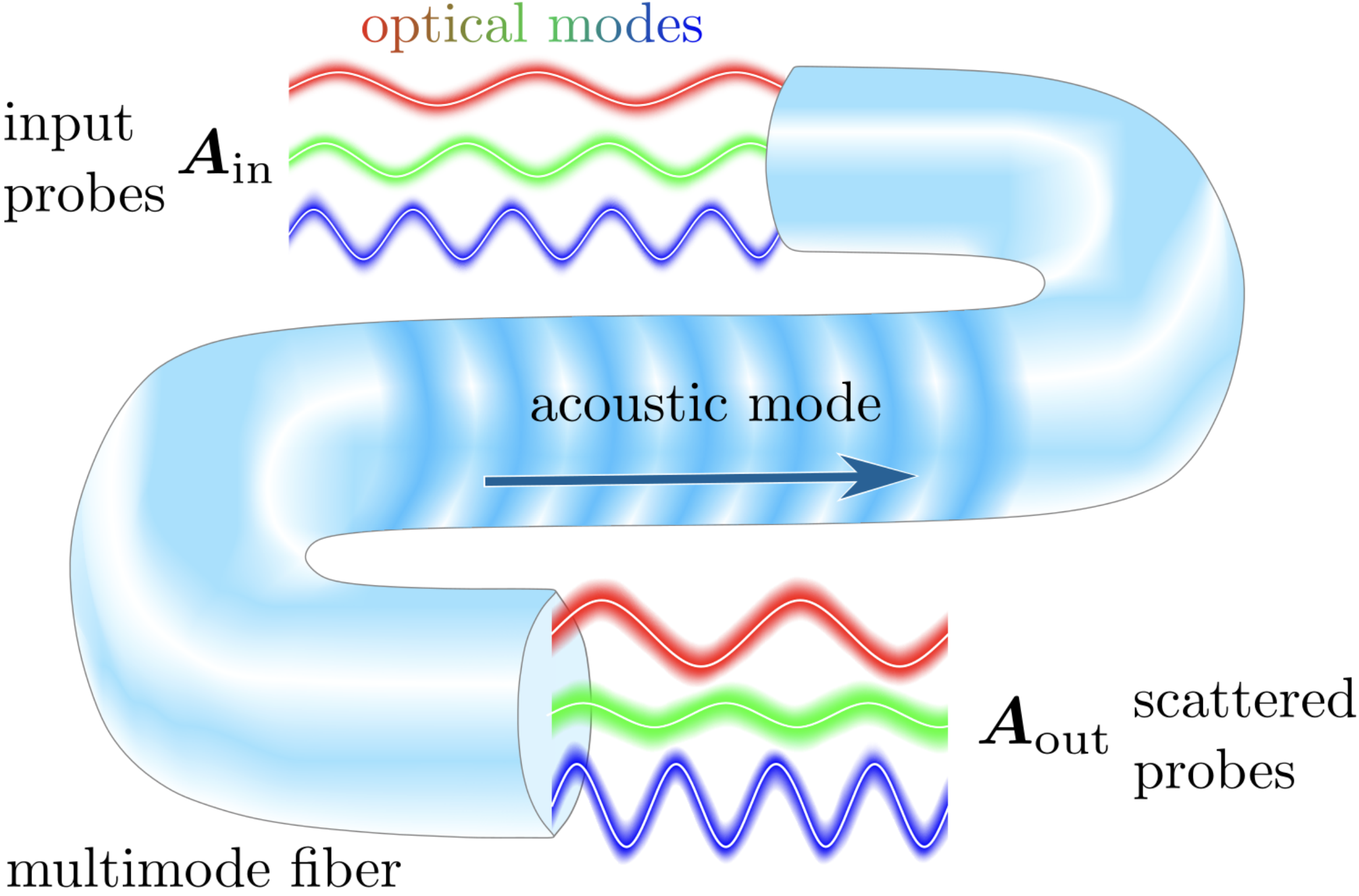}
    \caption{{\bf Schematic setup for off-resonant multimode stimulated Brillouin scattering.} Multiple frequency-detuned optical input probes interact via a single acoustic mode induced by multiple optical pumps (not depicted). As long as the probe frequency differences is on the order of the width of the Stokes peak, the input probes $\bm{A}_\text{in}$ and the scattered probes $\bm{A}_\text{out}$ are related by a non-Hermitian matrix $T$ with $\bm{A}_\text{out}=T\bm{A}_\text{in}$. The elements of $T$ can be tuned by changing pump amplitudes, pump phases and frequency spacing of both probes and pumps.}
    \label{fig:setup}
\end{figure}

Recently, the experimental demonstration of EP2s in an optical fiber opened a new avenue to explore exceptional points in a continuum optoacoustic platform~\cite{Bergman2021}.
The approach is based on stimulated Brillouin scattering (SBS), a coherent nonlinear optical effect that couples optical modes via acoustic modes in optical fibers -- well-known for its disturbing effect for telecommunications~\cite{Ippen1972}.
SBS has been experimentally studied for sensing and spectroscopy applications~\cite{ niklesSimpleDistributedFiber1996, hasegawaMeasurementBrillouinGain1999, geilenExtremeThermodynamicsNanolitre2023}, microwave photonics~\cite{ marpaungIntegratedMicrowavePhotonics2019}, biomedical imaging~\cite{Kabakova2024,Prevedel2019}, narrow-linewidth lasers~\cite{otterstromSiliconBrillouinLaser2018a, Gundavarapu2019, zengOpticalVortexBrillouin2023}, and recently, for optical memory~\cite{ zhuStoredLightOptical2007, geilenHighSpeedCoherentPhotonic2024, merkleinChipintegratedCoherentPhotonicphononic2017a, safferBrillouinbasedStorageQPSK2024, stillerCoherentlyRefreshingHypersonic2020} as well as neural network implementations~\cite{ beckerOptoacousticFieldprogrammablePerceptron2024,slinkovAllopticalNonlinearActivation2025}. 
The efficient coupling of optical and acoustic waves can be achieved in optical fibers and on-chip configurations \cite{ stillerPhotonicCrystalFiber2010, godetBrillouinSpectroscopyOptical2017, eggletonBrillouinIntegratedPhotonics2019, kittlausLargeBrillouinAmplification2016, kobyakovStimulatedBrillouinScattering2010a, neijtsOnchipStimulatedBrillouin2024, rodriguesCrossPolarizedStimulatedBrillouin2025,Ye2025, xuStrongOptomechanicalInteractions2023}, but is also established in bulk materials \cite{Chu2017,renningerBulkCrystallineOptomechanics2018a,PhysRevResearch.5.043140}. 
Largely studied in the classical domain, recent advances show experimental and theoretical progress towards the quantum realm \cite{blazquezmartinezOptoacousticCoolingTraveling2024a, cryer-jenkinsBrillouinMandelstamScattering2025b, enzianObservationBrillouinOptomechanical2019b, PhysRevLett.126.033601,zhuOptoacousticEntanglementContinuous2024, zhangQuantumCoherentControl2023b, StrongCoupling2025}.

In this work, we go beyond the demonstrated EP2s~\cite{Bergman2021}, and develop an off-resonant, multimode theory for SBS, where we utilize the fabrication-free nature of this platform allowing us to freely design mode structures.
The couplings between the modes are the result of the SBS process called Stokes scattering: in the simplest case, an optical probe and an optical pump counter-propagate in a fiber and interact via an acoustic mode close to resonance, approximately obeying energy conservation. 
We extend this picture to handle arbitrarily many pumps and probes off-resonantly interacting via a single acoustic mode, enabling symmetry-induced exceptional points of any order.
A schematic of this process is depicted in Fig.~\ref{fig:setup}.
Beyond the realization of such exceptional points, the developed off-resonant, multimode theory for SBS may find applications in multimode optoacoustic sensing, multi-frequency optoacoustic signal processing and synthetic neuromorphic computing.

\section{Results}

\subsection{Off-resonant multimode stimulated Brillouin scattering}
Multiple optical modes interacting via one or more acoustic modes result in a set of coupled differential equations~\cite{Wolff2021}.
Within the continuous-wave approximation, the resonant multimode evolution decouples and thus reduces the multimode dynamics to resonant single-probe-single-pump SBS.
Beyond resonant SBS, off-resonant SBS has been described in the one-probe-one-pump context~\cite{BoydBook} and for a specific two-probe-two-pump setup~\cite{Bergman2021}. The latter yields coupled dynamical equations for the evolution of the probe modes, which cannot be reduced to single-probe-single-pump dynamics.
Here, we generalize off-resonant SBS to arbitrary $N$-probe-$M$-pump setups encapsulating these previous results.
Starting from the off-resonant coupling of the optical modes to the acoustic mode we provide a complete derivation of the dynamical matrix governing the transmission of the probe amplitudes, as well as the physical interpretation of the process.

A configuration of $N$ optical probes and~$M$ optical pumps, which counterpropagate in a fiber, are coupled via acoustic modes, that can be approximated as a single acoustic mode if the probe and pump frequencies are within the typical Brillouin linewidth (around \SI{30}{MHz}), respectively.
The evolution of the probe modes is described by a dynamical matrix $H$ determined by the mode configuration and the complex pump-mode amplitudes.
All modes are described within the continuous-wave approximation propagating along the $z$-direction.
The optoacoustic interaction in the fiber is governed by a paraxial equation, for which the radial dependence of the optical and acoustic amplitudes decouples from the propagation. 
The probes (sometimes referred to as seeds) $\bm{E}_i^\textrm{s}$ and pumps $\bm{E}_m^\textrm{p}$ are described by the linearly polarized, radially integrated electric fields
\begin{subequations}
\label{eq:fields}
\begin{align}
    \bm{E}_i^\ups (z,t) &= A_i^\ups (z) \; e^{\ii(k_i^\ups z - \omega_i^\ups t)} \; \hat{\bm{e}} \equiv   A_i^\ups (z) \; e^{\ii\phi_i^\ups} \; \hat{\bm{e}}, \\
    \bm{E}_m^\upp (z,t) &= A_m^\upp \;e^{\ii(-k_m^\upp z - \omega_m^\upp t)} \; \hat{\bm{e}} \equiv   A_m^\upp \; e^{\ii\phi_m^\upp} \; \hat{\bm{e}} \, ,
\end{align}
\end{subequations}
where the subscripts $i\in\{1, \ldots,N\}$ ($m\in\{1,\ldots,M\}$) distinguishes different probes (pumps), $k$ and $\omega$ are the wavenumbers and frequencies of the modes, respectively, and $\hat{\bm{e}}$ is the polarization vector.
Note the $z$-independence of the pump amplitude $A_m^\upp$ as we work in the undepleted regime, i.e., in the limit of large pump powers $A_m^\text{p} \gg A_i^\text{s}$.
We assume that SBS induces a single acoustic mode $\Phi(z,t)$ propagating in the same direction as the pumps described by
\begin{equation}
    \Phi(z,t) = b(z) e^{\ii(-qz-\Omega t)}
\end{equation}
with amplitude $b(z)$, acoustic wavenumber $q$, and acoustic frequency $\Omega$.
Accounting for all off-resonant interactions within the continuous-wave approximation, the profile of probe amplitudes $A_i^\ups(z)$ and the acoustic mode amplitude $b(z)$ are governed by the following coupled equations derived in the Methods:
\begin{multline}
    \ii \left( \partial_z + \frac{\Gamma_i}{v_i}  \right) A_i^\ups(z)
    = -\frac{1}{\mathcal{P}_i} \sum_{\text{w}} \sum_m \bigl[ e^{\ii(\phi_m^\text{w}-\phi_i^\ups)} e^{\ii(qz+\Omega t)} \\
    \times Q^* b^*(z) \, \omega_m^\text{w} A_m^\text{w}(z) \bigr], \label{eq:diff_A}
\end{multline}
and
\begin{multline}
    b(z) = \ii\frac{2\Omega}{\mathcal{E}_b\Gamma_b} \sum_{\text{x},\text{y}}\sum_{n,j} \biggl[ e^{-\ii (\phi_j^\text{y}-\phi^\text{x}_n)} e^{\ii(qz+\Omega t)}\, \\[-0.5em] \times \frac{Q^* A_n^\text{x}(z) (A_j^\text{y}(z))^*}{1-\ii \Gamma^\text{xy}_{nj}}\biggr]  \, , \label{eq:diff_b}
\end{multline}
where $\text{x,y,w}\in \{\ups, \upp\}$, and $j,m,n\in\{1,\ldots, N\}$ \mbox{($\in\{1,\ldots, M\}$)} if $\text{x,y,w} = \ups$ ($\text{x,y,w}= \upp$).
Here, $\Gamma_i$ describes the linear loss of the optical mode, $v_i$ is the group velocity of the optical mode, $\mathcal{P}_i$ is the power density of the optical mode along the fiber, $\mathcal{E}_b$ is the energy density of the acoustic field, $\Gamma_b$ is the width of the Stokes peak of the optoacoustic interaction, $Q$ is the optomechanical overlap, which we assume to be mode independent, and the star denotes complex conjugation.
The degree of off-resonance is captured by
\begin{equation}
\Gamma^\text{xy}_{nj} = \frac{\Omega^2 - (\omega_n^\text{x}-\omega_j^\text{y})^2}{\Gamma_b \Omega}, \label{eq:gamma_def}
\end{equation}
which vanishes for resonant interactions $|\omega_n^\text{x}-\omega_j^\text{y}|=\Omega$~\cite{BoydBook}.
Inserting $b^*(z)$ into Eq.~\eqref{eq:diff_A} results in coupled first-order differential equations in terms of the optical mode amplitudes only.
We assume that linear losses of the probe amplitudes $\Gamma_i$ are negligible, which is reasonable for standard single-mode fibers, resulting in
\begin{multline}
    \label{eq:intermediate}
    \ii \partial_z A_i^\ups(z) = \ii\frac{2\Omega|Q|^2}{\mathcal{P}_i\mathcal{E}_b\Gamma_b} \sum_\text{w,x,y} \sum_{j,m,n} \Bigr[e^{\ii[(\phi_m^\text{w}-\phi_i^\ups)-(\phi^\text{x}_n-\phi_j^\text{y})]} \\
    \times \omega_m^\text{w} \frac{A_m^\text{w}(z) (A_n^\text{x}(z))^*}{1+\ii \Gamma^\text{xy}_{nj}} A_{j}^\text{y}(z) \Bigr] \, .
\end{multline}
Imposing a strict rotating-wave approximation on the evolution of the probe amplitudes yields
\begin{equation}\label{eq:freq_matching}
    (\omega_m^\text{w} -\omega_i^\ups) - (\omega_n^\text{x} - \omega_j^\text{y}) = 0 \, .
\end{equation}
Further probe-probe or pump-pump Brillouin interaction captured by $\text{x}=\text{y}$ can be neglected, because of the strong off-resonant damping $\Gamma_{nj}^\text{ss},\Gamma_{nj}^\text{pp},\gg1$ in the limit $\Omega\gg\Gamma_b$.
Thus, the probe and pump distinction is fixed to $\text{y}=\ups$ and $\text{w}=\text{x}=\upp$.
This simplifies Eq.~\eqref{eq:intermediate} to 
\begin{multline}
    \ii \partial_z A_i^\ups(z) = \ii\frac{2\Omega|Q|^2}{\mathcal{P}_0\mathcal{E}_b\Gamma_b} \sum_{j,m,n}^{\omega_m^\upp -\omega_i^\ups=\omega_n^\upp - \omega_j^\ups} \!\!\!\!\!\!\! \Bigr[e^{\ii z[(k_m^\upp+k_i^\ups)-(k^\upp_n+k_j^\ups)]} \\[-0.5em]
    \times \omega^\upp_m \frac{A_m^\upp (A_n^\upp)^*}{1+\ii \Gamma^\text{ps}_{nj}} A_{j}^\ups(z) \Bigr] \, , \label{eq:omega_matched}
\end{multline}
where we used Eq.~\eqref{eq:fields}.
We further assume that the power density of the probe modes is probe-independent, i.e., \mbox{$\mathcal{P}_i=\mathcal{P}_0$}.
Additionally, we recognize that the factor $\omega^\upp_m$ results in negligible changes of the effective Brillouin gain, $g_m=(2\omega^\upp_m\Omega|Q|^2)/(\mathcal{P}_0^2\mathcal{E}_b\Gamma_b)$, compared to the average Brillouin gain, $g_0=(2\omega_0^\upp\Omega|Q|^2)/(\mathcal{P}_0^2\mathcal{E}_b\Gamma_b)$, with the average pump frequency $\omega_0^\upp$. By substituting $A_q^\ups (z)=e^{-2ik_q^\ups z} a_q^\ups (z)$, we find 
\begin{multline}
    (2k_i^\ups+\ii \partial_z) a_i^\ups (z) = \ii g_0 \mathcal{P}_0 \!\!\!\!\!\!\! \sum_{j,m,n}^{\omega_m^\upp -\omega_i^\ups=\omega_n^\upp - \omega_j^\ups} \!\!\!\!\!\!\! \Bigr[e^{-\ii z[(k_m^\upp-k_i^\ups)-(k^\upp_n-k_j^\ups)]}
    \\[-0.5em] \times  \frac{A_m^\upp (A_n^\upp)^*}{1+\ii \Gamma^\text{ps}_{nj}} a_{j}^\ups(z) \Bigr] \, . \label{eq:k_matched}
\end{multline}
Assuming linear dispersion, i.e., $k \propto \omega$, makes the $z$-dependent phase factor vanish.
This also shows that our previous assumptions and approximation lead to only phase-matched interactions~\cite{Wolff2021}.
Finally, Eq.~\eqref{eq:k_matched} can be rewritten as the Schr\"{o}dinger-like equation $\ii \partial_z \bm{a}(z) = H \bm{a}(z)$
with $\bm{a}(z)=(a_1^\ups(z), \ldots , a_N^\ups(z))^T$ and dynamical matrix
\begin{align}
    H_{ij} = &-2k_i^\ups \delta_{ij} + \ii g_0 \mathcal{P}_0 \!\!\!\!\!\!\! \sum^{\omega_m^\upp -\omega_i^\ups=\omega_n^\upp - \omega_j^\ups}_{m,n} \frac{A_m^\upp (A_n^\upp)^*}{1+\ii \Gamma^\text{ps}_{nj}}. \label{eq:explicit_dyn_matrix}
\end{align}
The dynamical matrix is $z$-independent, and therefore, the differential equation for $\bm{a}(z)$ is formally solved by $\bm{a}(L) = T \bm{a}(0)$, where the transmission matrix $T$ is defined as
\begin{equation}\label{eq:transmission}
    T=e^{-\ii HL} \, ,
\end{equation}
and $L$ is the length of the fiber.
This transmission matrix governs the probe evolution depicted in Fig.~\ref{fig:setup}.

\begin{figure}[t]
    \centering
    \includegraphics[width=0.8\columnwidth]{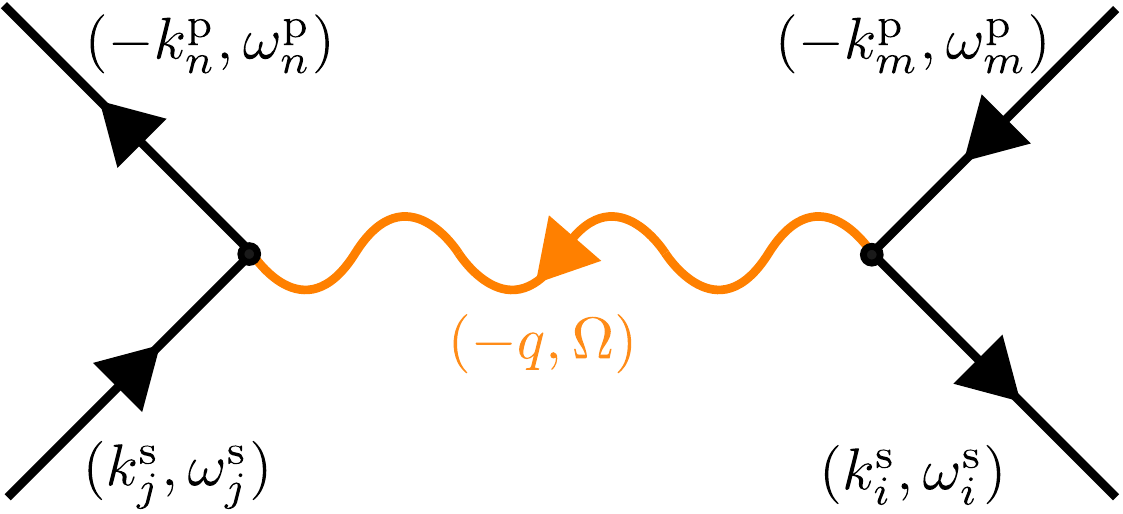}
    \caption{{\bf Coupling of probe modes interpreted as phonon-mediated photon-photon scattering.} The black lines depict the photons that interact via a phonon (orange line). The arrows indicate the propagation direction.}
    \label{fig:Feynman}
\end{figure}

Even though the dynamical matrix $H$ is derived in the continuous-wave, steady-state regime, it can be interpreted in the picture of phonon-mediated photon-photon scattering.
A probe and a pump photon interact with one another via a phonon in a process as sketched in Fig.~\ref{fig:Feynman}.
The phonon carries energy and momentum, and violates energy conservation during the scattering process, which results in the Lorentzian resonance peak stemming from the factor $1/(1-\ii \Gamma^\text{xy}_{nj})$ with $\Gamma^\text{xy}_{nj}$ in Eq.~\eqref{eq:gamma_def}.
However, energy-momentum conservation of the entire scattering process is ensured by the photon frequency matching, cf. Eq.~\eqref{eq:freq_matching}.
Therefore, the photon-photon scattering is elastic, and all inelastic scattering channels do not contribute to the multimode SBS within the approximations we make.

\subsection{Geometric representation of multimode stimulated Brillouin scattering}

\begin{figure}[ht]
    \centering
    \includegraphics[width=0.9\columnwidth]{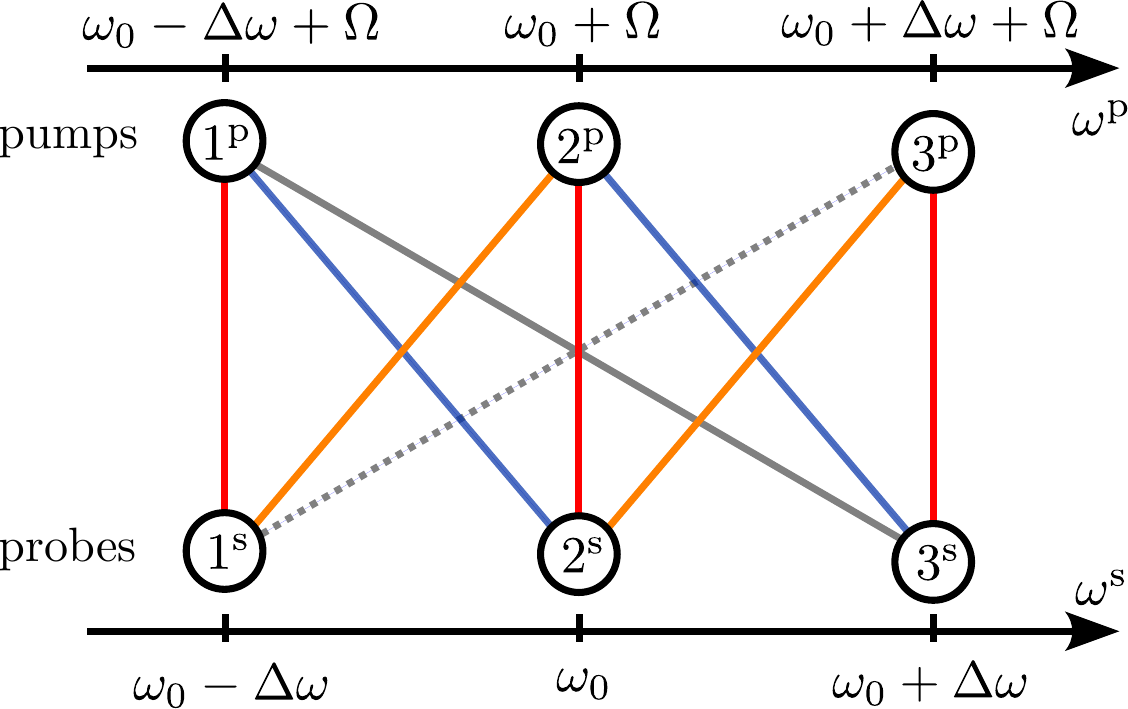}
    \caption{{\bf Geometric representation of an evenly spaced 3-probe-3-pump configuration.} The corresponding dynamical matrix~$H_3$ is given in Eq.~\eqref{eq:33effH}. Parallel lines are colored identically. The red, orange and blue lines appear more than once, indicating the appearance of off-diagonal terms in~$H_3$. The gray solid and dashed lines appear once, thus they effect the diagonal of $H_3$ only.}
    \label{fig:graph}
\end{figure}

Even though the dynamical matrix can be derived from Eq.~\eqref{eq:explicit_dyn_matrix}, we here present an alternative method to determine its form through a geometric representation.
For the geometric construction we associate a node with each probe and each pump.
All $N$ probes are arranged horizontally with the spacing according to the frequency $\omega_i^\ups$ of the respective mode.
Parallel to the array of probe-nodes, all~$M$ pumps are arranged horizontally according to their frequencies $\omega_n^\upp$.
The frequencies of the pumps are shifted by the acoustic-mode frequency $\Omega$ such that resonant probe-pump pairs with $\omega_n^\upp-\omega_i^\ups=\Omega$ are vertically aligned.
Next, lines indicating the interactions are drawn between all probe-pump pairs, and colored according to their orientation.
All parallel lines have the same color in the geometric representation.
As an example, we show an evenly spaced 3-probe-3-pump configuration in Fig.~\ref{fig:graph}.

From this geometric representation the dynamical matrix $H$ of size  $N\times N$ can be read off.
The diagonal elements $H_{jj}$ are given by
\begin{equation}
    H_{jj} = -2k_j^\ups+ \ii g_0 \mathcal{P}_0 \sum_{n=1}^M \frac{|A_n^\upp|^2}{1+\ii \Gamma_{nj}^\text{ps}}\, ,
\end{equation}
where the sum runs over all lines connecting to probe node $j$.
The additional real term $-2k_j^\ups$, not explicitly in the geometric representation, describes the trivial phase evolution of probe $a_j^\ups$.
For any off-diagonal element $H_{ij}$ we turn to the remaining lines.
We find a non-zero matrix element between two probes $i$ and $j$ if and only if the respective probe nodes are connected to any pump nodes by lines of the same color, e.g., all probe nodes in Fig.~\ref{fig:graph} are coupled to each other.
To obtain the off-diagonal matrix element we must sum over all identically colored lines $\{\mathcal{C}_{ij}\}$ connecting to node $i$ and $j$.
The matrix element that couples probe $i$ and $j$ is
\begin{equation}
    H_{ij} = \ii g_0 \mathcal{P}_0 \sum_{c\in\{\mathcal{C}_{ij}\}} \frac{ A_{\hat{n}(i,c)}^\upp (A_{\hat{n}(j,c)}^\upp)^*}{1+\ii \Gamma_{\hat{n}(j,c)j}^\text{ps}}\, ,
\end{equation}
where the map $\hat{n}(j,c)$ gives the index of the pump which is reached from the probe $j$ by the line of the color $c$.

To summarize, if the graphical representation for a specific experiment has been constructed we simply read off the dynamical matrix by considering the lines that are connected to the probe nodes. 
The color of the lines is used to determine the off-diagonal elements of the dynamical matrix~$H$.
This method for obtaining $H$ is especially useful for setups with multiple pumps and probes as it provides a simple and straightforward strategy to keep track of all couplings.
\vspace{1cm}
\subsection{Fabrication-free third-order exceptional points}

After working out off-resonant multimode SBS to derive the optically controlled dynamical matrix $H$, we come back to the initial question of how to generate exceptional points in such systems. We first focus on EP3s induced by anti-$\mathcal{PT}$ symmetry in a 3-probe-3-pump experiment with equal frequency spacing, expanding the results presented in Ref.~\citenum{Bergman2021}.
The probe frequencies are given by $\omega_i^\ups=\omega_0 + (i^\ups - 2)\Delta \omega$ with $i^\ups=1,2,3$ and the pump frequencies by $\omega_n^\upp=\omega_0 + (n^\upp - 2)\Delta \omega+\Omega$) with $n^\upp=1,2,3$, which results in the graphical representation shown in Fig.~\ref{fig:graph}.
We refer to such situations, where the probes, as well as the pumps, are equidistantly spaced in frequency, and separated by $\Omega$, as symmetric, and we consider the general symmetric $N$-probe-$N$-pump case below.
The dynamical matrix $H_3$ of such a symmetric 3-pump-3-probe setup is given by
\begin{widetext}
    \begin{equation}\label{eq:33effH}
        H_3 = \ii g_0 \mathcal{P}_0 \begin{pmatrix}
         |A_1^\upp|^2 + \frac{|A_2^\upp|^2}{1+\ii \gamma}+ \frac{|A_3^\upp|^2}{1+2 \ii \gamma} - \frac{\ii \Gamma_b \gamma}{c g_0 \mathcal{P}_0} & A_1^\upp (A_2^\upp)^* + \frac{A_2^\upp (A_3^\upp)^*}{1+\ii \gamma} & A_1^\upp (A_3^\upp)^* \\
        A_2^\upp (A_1^\upp)^* + \frac{A_3^\upp (A_2^\upp)^*}{1+\ii \gamma} & \frac{|A_1^\upp|^2}{1-\ii \gamma} + |A_2^\upp|^2 + \frac{|A_3^\upp|^2}{1+\ii \gamma} & A_2^\upp (A_3^\upp)^* + \frac{A_1^\upp (A_2^\upp)^*}{1-\ii \gamma} \\
        A_3^\upp (A_1^\upp)^* & A_3^\upp (A_2^\upp)^* + \frac{A_2^\upp (A_1^\upp)^*}{1-\ii \gamma}  & \frac{|A_1^\upp|^2}{1-2\ii \gamma} + \frac{|A_2^\upp|^2}{1-\ii \gamma} + |A_3^\upp|^2 + \frac{\ii \Gamma_b \gamma}{c g_0 \mathcal{P}_0}
    \end{pmatrix}  - 2k_0 \, \mathbb{1}_3 \, ,
    \end{equation}
\end{widetext}
where $\Gamma^\text{ps}_{nj}$ is linearized in $\Delta \omega$ as $\Gamma_b\gg\Delta\omega$, the normalized detuning $\gamma$ is given by $\gamma=2\Delta\omega/\Gamma_b$, and $c$ denotes the speed of light in the fiber.
Without further constraints, the dynamical matrix $H_3$ has seven free parameters: setting $A_n^\upp = |A_n^\upp|e^{\ii \phi_n^\upp}$ with $n=1,2,3$, there are the three pump amplitudes $|A_n^\upp|$, the three pump phases $\phi_n^\upp$, and the frequency spacing $\Delta\omega>0$.
With this number of free parameters EP3s can generically be realized, because the $2(N-1)=4$ constraints can be generically met in four or more dimensions \cite{Sayyad2022}.
By fixing all pump phases~$\phi_n^\upp$ constant and tuning the three pump amplitudes and the frequency detuning we can realize an EP3 surrounded by a trefoil knot structure of EP2s, as experimentally realized in a cavity optomechanical system in Ref.~\citenum{Patil2022}.
However, this parameter space is vast and the EP3 is difficult to find experimentally, as its position depends on the fixed pump phases~$\phi_n^\upp$ and one would need to measure the full four-dimensional parameter space densely.
Instead, we reduce the codimension of the exceptional points by imposing anti-$\mathcal{PT}$ symmetry on the dynamical matrix, which reduces the number of constraints to two~\cite{Montag2024, Montag2024_2}.
We define the traceless part of $H_3$ as $\Tilde{H}_3=H_3-\left[\tr(H_3)/3\right]\mathbb{1}_3$ and constrain the pump phases as $\phi^\upp_1-2\phi^\upp_2+\phi^\upp_3=0$.
Then, $\Tilde{H}_3$ is anti-$\mathcal{PT}$ symmetric,
\begin{equation}
    \Tilde{H}_3 = - \Theta_3\Tilde{H}_3^*\Theta_3^{-1},
\end{equation}
where~$\Theta_3$ is a unitary matrix~\cite{Montag2024_2}, given in our case by the exchange matrix
\begin{equation}
    \Theta_3 = \begin{pmatrix}
        0&0&1\\
        0&1&0\\
        1&0&0
    \end{pmatrix} \, .
\end{equation}
This reflects the mirror symmetry of probes and pumps, see Fig.~\ref{fig:graph}, by exchanging $i^\ups=1$ with $i^\ups=3$ as well as $n^\upp=1$ with $n^\upp=3$, respectively.

\begin{figure}
    \centering
    \includegraphics[width=\columnwidth]{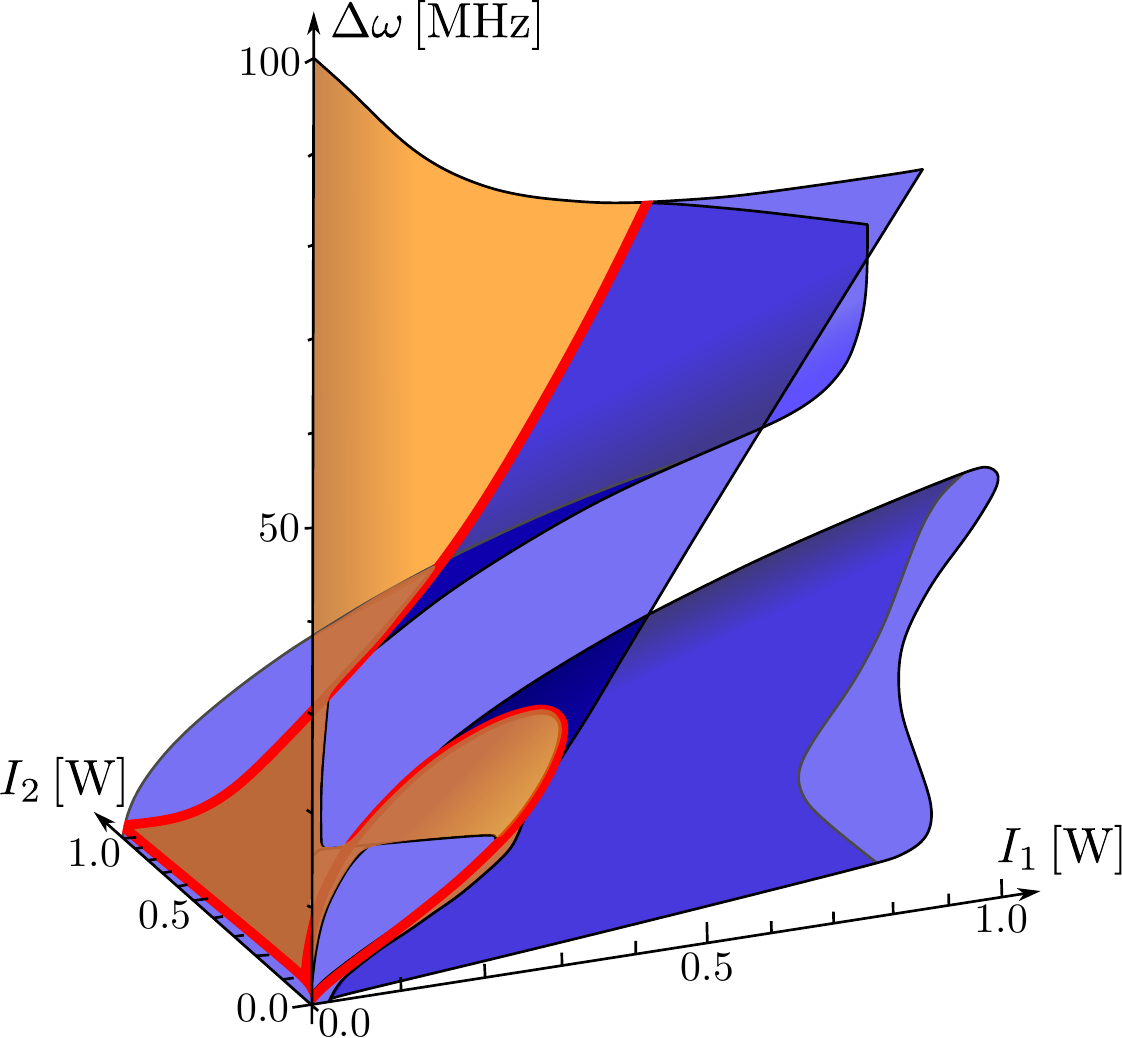}
    \caption{{\bf Spectral features of the dynamical matrix in parameter space.} EP3 lines (red), EP2 surfaces (blue) and the three-level imaginary Fermi surface (orange) in the three-dimensional parameter space spanned by $I_1$, $I_2$, and $\Delta \omega$. The EP3 line marks the fold line of the EP2 surface, and also the edge of the three-level imaginary Fermi surface.
    Here, $g_0=\SI{1.25}{(Wm)^{-1}}$, $\Gamma_b= \SI{45.6}{MHz}$, and $c=c_0/1.44$ with $c_0$ the speed of light in vacuum.}
    \label{fig:3D_parameter_space}
\end{figure}

Given this symmetry, we can use the results from Ref.~\citenum{Montag2024} to show the emergence of lines of EP3s in the experimentally accessible three-dimensional parameter space spanned by $\gamma,I_1,I_2>0$, where the intensities $I_\alpha$ are given by $I_\alpha=\mathcal{P}_0 |A_\alpha^\upp|^2$.
We define
\begin{equation}
    \nu = \frac{\det[\Tilde{H}_3]}{2} \quad \text{and} \quad \eta = -\frac{\tr[(\Tilde{H}_3)^2]}{6} \, ,
\end{equation}
with \mbox{$\nu\in \ii \mathbb{R}$} and $\eta\in\mathbb{R}$ due to the imposed anti-$\mathcal{PT}-$symmetry.~\cite{Sayyad2022,Montag2024,Montag2024_2}.
The EP3s emerge along the line defined by $\nu=\eta=0$.
However, due to the presence of anti-$\mathcal{PT}$ symmetry, these lines of EP3s do not appear isolated but are part of higher-dimensional spectral structures~\cite{Montag2024,Montag2025}.
The line of EP3s is both the fold line of the EP2 surface defined by $\nu^2+\eta^3=0$ and the edge of a three-level imaginary bulk Fermi surface~\cite{Kozii2017,Zhou2018}, on which all eigenvalues have the same imaginary part with distinct real parts, given by $\nu=0$ for $\eta<0$.
We exemplify all spectral features in Fig~\ref{fig:3D_parameter_space}.

In the three-dimensional parameter space, the EP3 line can be parametrically encircled and the measured eigenvalues exchange in unique patterns.
Importantly, both the EP2 and imaginary Fermi surfaces can be detected, and be used as guides towards the EP3 line.
This, together with the low codimension of the EP3s, is a major advantage of imposing anti-$\mathcal{PT}$ symmetry.

\subsection{Exceptional points in transmission matrix}

What has not been discussed yet is how to measure exceptional points in the proposed setup, where we cannot measure the dynamical matrix directly but only have access to the transmission matrix according Eq.~\eqref{eq:transmission}.
Let us define $T_3 = \mathrm{exp}(-\ii H_3 L)$ as the transmission matrix associated with the symmetric 3-probe-3-pump setup.
It can be determined by comparing ingoing and outgoing probe modes, for which both amplitude and relative phase are measured using homodyne measurements.

As $T_3$ is the matrix exponential of the non-Hermitian matrix $G\equiv-\ii H_3 L$, it (i) is also non-Hermitian, (ii) has non-zero complex eigenvalues, and (iii) exhibits exceptional points of the same order as $H$ at the same points in parameter space.
Mapping the imaginary Fermi surface of $H_3$ to $T_3$ is slightly more subtle. 
It corresponds to a real Fermi surface of $G$, where all eigenvalues are still non-degenerate, but all real parts are identical.
Therefore, the eigenvalues of $T_3$ are distributed on a circle in the complex plane, corresponding to eigenvalues of equal magnitude.
Physically, these magnitudes correspond to the Brillouin gain, and thus, the Fermi arcs are directly observable as equal amplification arcs.

\begin{figure*}
    \centering
    \includegraphics[width=1.0\linewidth]{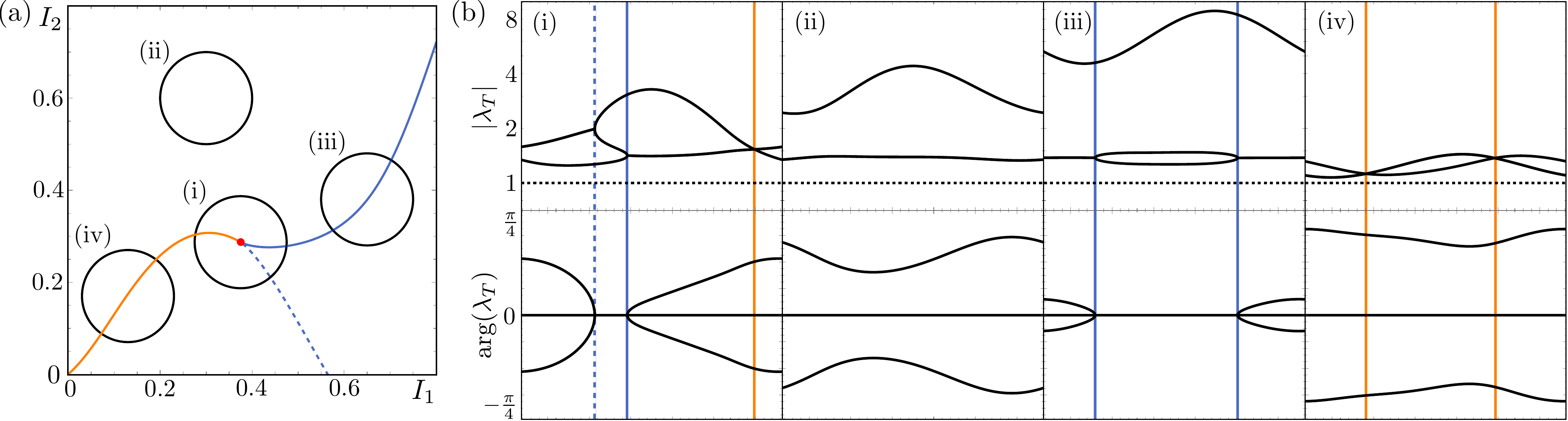}
    \caption{{\bf Detection of EP3s by scanning along closed loops in parameter space.} (a) Spectral features in the two-dimensional parameter space spanned by the pump laser intensities $I_1$ and $I_2$ for constant frequency detuning $\Delta \omega = \SI{60}{MHz}$, with all other parameters as in Fig.~\ref{fig:3D_parameter_space}. The symmetry-induced EP3 is marked by a red dot, the second-order exceptional lines are shown in blue and the three-level imaginary Fermi arc is plotted in orange. Four distinct parameter loops are indicated. (b) Absolute value and argument of the transmission matrix eigenvalues along the closed loops shown in (a). Position of special spectral features of the dynamical matrix are indicated by vertical lines. The horizontal dashed line in the upper row marks the onset of gain due to Brillouin scattering.}
    \label{fig:2D_parameter_space}
\end{figure*}

Given the rather vertical orientation of the exceptional line at larger frequency spacing, observable in Fig.~\ref{fig:3D_parameter_space}, it is easiest to detect the exceptional line by measuring along closed loops on horizontal cross sections of the parameter space.
This is achieved by fixing the frequency spacing~$\Delta \omega$ and then varying the pump intensities $I_1$ and $I_2$.
The reduced parameter space is a plane, cf. Fig.~\hyperref[fig:2D_parameter_space]{5(a)}, on which the third-order exceptional lines are reduced to points (red) and the second-order exceptional surfaces and the imaginary Fermi surface appear as lines (blue and orange, respectively).
Measuring the eigenvalues of the transmission matrix along the closed curves, indicated in Fig.~\hyperref[fig:2D_parameter_space]{5(a)}, results in the curves shown in Fig.~\hyperref[fig:2D_parameter_space]{5(b)}.
The eigenvalues $\lambda_T$ are plotted there in terms of magnitude and phase, as discussed above.
By comparing panels in Fig.~\hyperref[fig:2D_parameter_space]{5(b)} it is clear how a closed loop encircling the EP3 can be distinguished from any other loop.
Importantly, we have to distinguish the two different possibilities for EP2s: the degeneracy can either occur at a larger (upper EP2, blue dashed) or a smaller (lower EP2, blue solid) magnitude with respect to the remaining eigenvalue.
If and only if the EP3 is encircled once, each spectral feature -- the lower second-order exceptional line, the upper second-order exceptional line, and the equal amplification line -- intersect the parameter loop once.
For any loop that does not encircle the EP3, each feature can only appear an even number of times if it appears at all, and it always appears twice in a row when parametrically tuning along the loop.
Therefore, we can determine the position of the EP3 by finding a loop that shows all three features once and subsequently constricting this loop down to a point, while keeping all three features.

\subsection{Symmetry-induced higher-order exceptional points}

The approach presented above can be generalized to exceptional points of any order $N>3$.
As before we consider a symmetric $N$-probe-$N$-pump setup, with probe frequencies $\omega_i^\ups=\omega_0 + (i^\ups - \frac{N+1}{2})\Delta \omega$ with $i^\ups\in\{1,\ldots,N\}$.
Each probe is in resonance with a pump, thus the pump frequencies are given by $\omega_n^\upp=\omega_0 + (n^\upp - \frac{N+1}{2})\Delta \omega+\Omega$ with $n^\upp\in\{1,\ldots,N\}$.
Constructing the associated graphical representation and deriving the dynamical matrix $H_N$ from it results in an $N \times N$ dynamical matrix $H_N$ with $2N+1$ free parameters, namely the frequency spacing $\Delta \omega$, $N$ pump amplitudes and $N$ pump phases.
While the number of parameters is large enough to realize EP$N$s, we find that the phases need to be tunable parameters in order to do so.
Because freely tuning the phases provides experimental challenges, and to reduce the size of the parameter space, we again make use of anti-$\mathcal{PT}$ symmetry.
It can be shown that by imposing
\begin{equation}\label{eq:sym}
    A_n^\upp = A_{N+1-n}^\upp \, ,
\end{equation}
the traceless part of the dynamical matrix, $\Tilde{H}_N=H_N-\left[\tr(H_N)/N\right]\mathbb{1}_N$, fulfills anti-$\mathcal{PT}$ symmetry, $\Tilde{H}_N=-\Theta_N\Tilde{H}_N^*\Theta_N^{-1}$, where the generator of the symmetry is again the exchange matrix
\begin{equation}
    \Theta_N = \begin{pNiceMatrix}
             0 & \Cdots &  0 & 1    \\
             \Vdots & \Iddots & \Iddots & 0 \\
             0 & \Iddots & \Iddots & \Vdots \\
             1 & 0 & \Cdots & 0
          \end{pNiceMatrix}_{N \times N} \, .
\end{equation}
Further, the constraints in Eq.~\eqref{eq:sym} reduce the number of tunable parameters to $2\lceil \frac{N}{2}\rceil+1$, where $\lceil.\rceil$ is the ceiling function, assuming that the amplitudes and phases of the pumps $i^\upp\in\{1,\ldots,\lceil \frac{N}{2}\rceil\}$ and the detuning are freely tunable.
Due to the reduced codimension of the EP$N$, the number of parameters is enough to tune the dynamical matrix toward it~\cite{Sayyad2022,Montag2024}.
In addition to the emergence of the exceptional points in the remaining lower-dimensional parameter space, the induced exceptional points will be again surrounded by intricate exceptional structures~\cite{Montag2025}.
These structures will translate to the transmission matrix of the $N$-probe-$N$-pump setup and are therefore accessible in experiment.
The exceptional structures will be accompanied by extended Fermi structures, which may be observed directly as signature of the transmission matrix as well.
This shows that off-resonant Brillouin scattering experiments are a potential platform for the realization of symmetry-induced exceptional points of any order.
However, note that even in presence of anti-$\mathcal{PT}$ symmetry, no exceptional points, apart from the EP3 discussed above, can be detected by simply measuring along closed loops.
The reason is the increase of the codimension of the exceptional points of order $N>3$, which requires measurements on higher-dimensional closed hypersurfaces.

\section{Discussion}

In this work we have developed a general formalism for the treatment of off-resonant multimode SBS in the undepleted regime.
Within this regime, the coupled nonlinear equations describing the optoacoutic interaction reduce to a set of coupled linear first-order differential equations for the probe modes, encoded by a dynamical matrix $H$.
The elements of $H$ are determined by the pump amplitudes and pump phases, as well as the frequencies of both probes and pumps.
To construct the dynamical matrix for an arbitrary setup of probes and pumps we have introduced a geometrical representation of the probe-pump setup, from which the elements of $H$ are uniquely determined.
The high degree of tunability of $H$ given the choice of the number of probes and pumps and their respective frequencies allows us to study the emergence of higher-order exceptional points in this platform, going beyond the previously found EP2s~\cite{Bergman2021}.

By imposing anti-$\mathcal{PT}$ symmetry on the dynamical matrix we have shown explicitly how third-order exceptional points arise in the spectrum of the dynamical matrix $H_3$, associated with a symmetric 3-probe-3-pump experiment.
Due to the presence of this symmetry, the number of parameters that need to be fine-tuned at the exceptional point is reduced to two.
Thus, the EP3s of $H_3$ arise as continuous lines in the experimentally accessible parameter space spanned by the frequency detuning and two pump intensities.
Besides the low codimension, an additional advantage of studying these symmetry-induced EP3s is the presence of extended spectral features such as second-order exceptional surfaces and so-called imaginary bulk Fermi surfaces, which both connect to the line of EP3s.
Finding these extended structures in parameter space is experimentally easier than finding the third-order exceptional line directly and thus, they can be used as guiding beacons towards the EP3s.
To enable this, we have discussed in detail how both the exceptional point and the three-level imaginary Fermi surfaces translate from the spectrum of the dynamical matrix to the eigenvalues of the transmission matrix.
Only the transmission matrix is measurable and this translation allows for the direct detection of the exceptional points, in the measurement of the probe dependent Brillouin gain.
Building on the results of the symmetric 3-probe-3-pump setup, a general framework for the generation of symmetry-induced exceptional points of any order in $N$-probe-$N$-pump setups has been derived.
This opens a new avenue for the fabrication-free generation of high-order exceptional points.
Beyond that, EPs of any order up to $N$ may appear in generic non-symmetric $N$-probe-$M$-pump setups with $N\neq M$, however, their occurrence must be shown in single-case studies.

Importantly, it has recently been noted that the weaker condition of having a similarity~\cite{Montag2024_2,Montag2025} already reduces the codimension of exceptional points.
To realize a similarity-induced EP3 in the context of this work, we impose the similarity called pseudo anti-Hermiticity~\cite{Montag2024_2}, 
$\Tilde{H}_3 = - \varsigma \Tilde{H}_3^\dagger\varsigma^{-1}$, where $\varsigma$ is an invertible matrix.
Compared to anti-$\mathcal{PT}$ symmetry, this is a weaker requirement as one requires unitarity for a symmetry generator.
The matrix~$\varsigma$ exists as long as $|A_1^\upp|=|A_3^\upp|$, and may depend on the pump amplitudes, the pump phases and the detuning.
Thus, imposing pseudo anti-Hermiticity effectively removes all constraints on the pump phases, while retaining an EP3 in the setup.
Note that in the symmetric 2-pump-2-probe case~\cite{Bergman2021}, pseudo anti-Hermiticity and anti-$\mathcal{PT}$ symmetry are equivalent requirements~\cite{Montag2024_2}.

Besides the highlighted generation of high-order exceptional points, the off-resonant multimode SBS derived here can be employed for the accurate description of a broad class of multimode experiments, such as distributed Brillouin sensing with multiple probes, multi-frequency optoacoustic signal processing, high-capacity memory for light based on traveling acoustic waves, microwave photonic filters and delay-lines, multimode quantum optomechanics and synthetic neuromorphic computing mediated by acoustic waves.
Another interesting application of the methods developed here is the study of the Riemann surfaces in the spectrum of the transmission matrix if only the pump phases are varied while keeping their intensities constant.
This results in a periodic parameter space, likewise to solid state systems, in which for example the braiding of exceptional points~\cite{PhysRevResearch.5.L042010, wang2021topological} and the appearance of non-contractible Fermi cuts~\cite{Montag2025b} may be studied.

\section*{Acknowledgments}
The authors would like to thank Grigorii Slinkov for insightful discussions. A.M., J.T.G., and F.K.K. acknowledge funding from the Max Planck Society Lise Meitner Excellence Program~\mbox{2.0}. A.M., J.T.G. and F.K.K. also acknowledge support from the European Union’s ERC Starting Grant “NTopQuant” (101116680). The views expressed are those of the authors and do not necessarily reflect those of the European Union or the ERC. Q.L. acknowledges support from the École Normale Supérieure Paris-Saclay while contributing to the work reported here. B.S. acknowledge funding from the  Max Planck Research Group Scheme and the DFG grant \mbox{STI-792/1-1.}

\section*{Author Contributions}

A.M. and J.T.G. developed the framework for off-resonant multimode SBS, based on initial work of Q.L. on the symmetric $3$-probe-$3$-pump case. Q.L. developed the geometric representation of the probe-pump configuration. A.M. carried out the non-Hermitian symmetry analysis, mapped the spectral features of the dynamical matrix to the transmission matrix, and proposed realizations of higher-order exceptional points. B.S. and F.K.K. initiated and supervised the project. A.M. prepared the manuscript with support of J.T.G. and input from all co-authors.

\section*{Data Availability}
Data sharing not applicable to this article as no datasets were generated or analyzed during the current study.

\section*{Methods}

\subsection*{Derivation of coupled mode equations}
We do not derive the coupled mode equations from the microscopic picture directly, but instead extend on established results on resonant multimode Brillouin scattering to account for off-resonant interactions.
To arrive at the coupled mode equation for resonant Brillouin scattering the following fundamental assumptions are made: (i) probes and pumps interact via a single acoustic mode, (ii) the interaction is governed by a paraxial equation, and (iii) probes and pumps are linearly polarized. This leads to Eqs.~(53a) and~(53b) in Ref.~\citenum{Wolff2021}, which we repeat here and adapt for our convenience.
The evolution of the $N$ probe and $M$ pump modes is given by
\begin{multline*}
    \ii \left(\partial_t + v_i\partial_z + \Gamma_i  \right) A_i^\ups(z)
    = -\frac{1}{\mathcal{E}_i} \sum_{\text{w}\in\text{p,s}} \sum_m \Bigl[ e^{\ii(\phi_m^\text{w}-\phi_i^\ups)} \\
    \begin{aligned}
        \times \Bigl(&\underbrace{e^{\ii(qz+\Omega t)} Q_{im}^* b^*(z) \, \omega_m^\text{w}}_{\text{Stokes}} \\
         &+ \underbrace{e^{-\ii (qz+\Omega t)} Q_{im} b(z) \, \omega_m^\text{w}}_{\text{anti-Stokes}}\Bigr) \, A_m^\text{w}(z) \Bigr] \, ,
    \end{aligned}
\end{multline*}
where the sum runs over all $\text{w}$ being probes and pumps, $m$ runs over $m \in \{1,\ldots,N\}$ ($m \in \{1,\ldots,M\}$) if $\text{w}=\ups$ ($\text{w}=\upp$), and we choose the probes to propagate in positive $z$-direction.
This is equivalent to Eq.~(53a) in \cite{Wolff2021}.
Because we consider continuous-wave operation of all optical modes, the time derivative vanishes.
The contributions in the sum are split into two processes, Stokes and anti-Stokes, and we only consider the contribution from the Stokes process in the main text.
With these simplifications we arrive at Eq.~\eqref{eq:diff_A} in the main text by dividing by $v_i$ and identifying $v_i\mathcal{E}_i=\mathcal{P}_i$.

To include off-resonant nature of the Brillouin scattering, we have to consider the contribution of each optical-mode pair to the acoustic mode evolution separately.
For a single pair of optical modes $A_n^\text{x}$ and $A_j^\text{y}$, we find for the contribution to the acoustic mode $b^\text{xy}_{nj}$ the dynamical equation
\begin{multline*}
    (\partial_t + v_b \partial _z + \Gamma_b) \, b^\text{xy}_{nj}(z)= \ii\frac{\Omega}{\mathcal{E}_b}e^{\ii(qz+\Omega t)}\\
    \begin{aligned}
        \times\Bigl[ &e^{-\ii (\phi_j^\text{y}-\phi^\text{x}_n)}
    Q_{nj}^* A_n^\text{x}(z) (A_j^\text{y}(z))^* \\
    &+ e^{\ii(\phi_j^\text{y}-\phi^\text{x}_n)} \, Q_{nj} (A_n^\text{x}(z))^* (A_j^\text{y}(z)) \Bigr] \, , 
    \end{aligned}
\end{multline*}
which is Eq.~(53b) in \cite{Wolff2021} adopted to accommodate off-resonant terms below.
Because we are considering the continuous wave regime, we drop the partial time derivative, and furthermore assume local acoustic response by setting $v_b=0$ due to $v_b \ll v_i$, i.e., the speed of sound is much smaller than the speed of light in the fiber.

To incorporate off-resonant Brillouin scattering we use methods from Chapter 9 in Ref.~\citenum{BoydBook} and substitute 
\begin{equation*}
    \Gamma_b \rightarrow \Gamma_b \, \left(1-\ii\frac{\Omega^2-(\omega^\text{x}_n-\omega^\text{y}_j)^2}{\Omega\Gamma_b} \right) \equiv \Gamma_b \, (1 - \ii \Gamma^\text{xy}_{nj}) \, ,
\end{equation*}
cf. Eq.~\eqref{eq:gamma_def}, which yields
\begin{multline*}
     b^\text{xy}_{nj}(z) = \ii\frac{\Omega }{\mathcal{E}_b\Gamma_b} e^{\ii(qz+\Omega t)}\\
     \begin{aligned}
         \times \biggl[&e^{-\ii (\phi_j^\text{y}-\phi^\text{x}_n)} \, \frac{Q_{nj}^* A_n^\text{x}(z) (A_j^\text{y}(z))^*}{1-\ii \Gamma^\text{xy}_{nj}}
     \\
     &+ e^{\ii(\phi_j^\text{y}-\phi^\text{x}_m)} \, \frac{Q_{nj} (A_n^\text{x}(z))^* (A_j^\text{y}(z))}{1-\ii \Gamma^\text{xy}_{nj}} \biggr] \, .
     \end{aligned}
\end{multline*}
The full acoustic mode profile is found by summing over all pairs of optical modes and summing up the individual contributions to the acoustic mode as $b(z)=\sum_\text{x,y} \sum_{n,j} b^\text{xy}_{nj}(z)$, i.e.,
\begin{multline*}
     b(z) = \ii\frac{\Omega }{\mathcal{E}_b\Gamma_b} e^{\ii(qz+\Omega t)}\\
     \begin{aligned}
         \times \sum_{\text{x},\text{y}}\sum_{n,j} \biggl[&e^{-\ii (\phi_j^\text{y}-\phi^\text{x}_n)} \, \frac{Q_{nj}^* A_n^\text{x}(z) (A_j^\text{y}(z))^*}{1-\ii \Gamma^\text{xy}_{nj}}
     \\
     &+ e^{\ii(\phi_j^\text{y}-\phi^\text{x}_m)} \, \frac{Q_{nj} (A_n^\text{x}(z))^* (A_j^\text{y}(z))}{1-\ii \Gamma^\text{xy}_{nj}} \biggr] \, .
     \end{aligned}
\end{multline*}
This can be simplified by relabeling $(\text{x},n)\leftrightarrow (\text{y},j)$ in the second term in the sum (third line in the equation above), and subsequently using $Q_{jn}^*=Q_{nj}$, which follows from the Hermiticity of the optomechanical overlap, and $\Gamma^\text{yx}_{jn}=\Gamma^\text{xy}_{nj}$ from the definition of the off-resonant coupling in Eq.~\eqref{eq:gamma_def}.
Assuming constant mode overlap $Q_{nj} \equiv Q$, justified by the small difference in the optical frequencies, yields Eq.~\eqref{eq:diff_b}.

\bibliography{references.bib}

@article{Bergman2021,
  title={Observation of anti-parity-time-symmetry, phase transitions and exceptional points in an optical fibre},
  author={Bergman, Arik and Duggan, Robert and Sharma, Kavita and Tur, Moshe and Zadok, Avi and Al{\`u}, Andrea},
  journal={Nature communications},
  volume={12},
  number={1},
  pages={486},
  year={2021},
  publisher={Nature Publishing Group UK London},
  doi={https://doi.org/10.1038/s41467-020-20797-7}
}

@article{Sayyad2022,
  title = {Realizing exceptional points of any order in the presence of symmetry},
  author = {Sayyad, Sharareh and Kunst, Flore K.},
  journal = {Phys. Rev. Res.},
  volume = {4},
  issue = {2},
  pages = {023130},
  numpages = {15},
  year = {2022},
  month = {May},
  publisher = {American Physical Society},
  doi = {10.1103/PhysRevResearch.4.023130},
  url = {https://link.aps.org/doi/10.1103/PhysRevResearch.4.023130}
}

@article{Patil2022,
  title={Measuring the knot of non-Hermitian degeneracies and non-commuting braids},
  author={Patil, Yogesh SS and H{\"o}ller, Judith and Henry, Parker A and Guria, Chitres and Zhang, Yiming and Jiang, Luyao and Kralj, Nenad and Read, Nicholas and Harris, Jack G E},
  journal={Nature},
  volume={607},
  number={7918},
  pages={271--275},
  year={2022},
  publisher={Nature Publishing Group UK London},
  url = {https://doi.org/10.1038/s41586-022-04796-w}
}

@article{Montag2024,
  title = {Symmetry-induced higher-order exceptional points in two dimensions},
  author = {Montag, Anton and Kunst, Flore K.},
  journal = {Phys. Rev. Res.},
  volume = {6},
  issue = {2},
  pages = {023205},
  numpages = {9},
  year = {2024},
  month = {May},
  publisher = {American Physical Society},
  doi = {10.1103/PhysRevResearch.6.023205},
  url = {https://link.aps.org/doi/10.1103/PhysRevResearch.6.023205}
}

@article{Montag2024_2,
  title={Essential implications of similarities in non-Hermitian systems},
  author={Montag, Anton and Kunst, Flore K},
  journal={Journal of Mathematical Physics},
  volume={65},
  number={12},
  year={2024},
  publisher={AIP Publishing},
  url={https://doi.org/10.1063/5.0206211}
}

@article{Montag2025,
  title = {Analytically tractable zoo of similarity-induced exceptional structures},
  author = {Montag, Anton and Isaacs, Jordan and St\aa{}lhammar, Marcus and Kunst, Flore K.},
  journal = {Phys. Rev. Res.},
  volume = {7},
  issue = {4},
  pages = {043199},
  numpages = {15},
  year = {2025},
  month = {Nov},
  publisher = {American Physical Society},
  doi = {10.1103/bzqh-2w3l},
  url = {https://link.aps.org/doi/10.1103/bzqh-2w3l}
}

@article{Montag2025b,
	title={{Spectral Riemann sheet topology of gapped non-Hermitian systems}},
	author={Anton Montag and Alexander Felski and Flore K. Kunst},
	journal={SciPost Phys.},
	volume={20},
	pages={133},
	year={2026},
	publisher={SciPost},
	doi={10.21468/SciPostPhys.20.5.133},
	url={https://scipost.org/10.21468/SciPostPhys.20.5.133},
}

@article{Delplace2021,
  title = {Symmetry-Protected Multifold Exceptional Points and Their Topological Characterization},
  author = {Delplace, Pierre and Yoshida, Tsuneya and Hatsugai, Yasuhiro},
  journal = {Phys. Rev. Lett.},
  volume = {127},
  issue = {18},
  pages = {186602},
  numpages = {6},
  year = {2021},
  month = {Oct},
  publisher = {American Physical Society},
  doi = {10.1103/PhysRevLett.127.186602},
  url = {https://link.aps.org/doi/10.1103/PhysRevLett.127.186602}
}

@article{Wolff2021,
	title = {Brillouin scattering—theory and experiment: tutorial},
	volume = {38},
	copyright = {© 2021 Optical Society of America},
	issn = {1520-8540},
	url = {https://opg.optica.org/josab/abstract.cfm?uri=josab-38-4-1243},
	doi = {10.1364/JOSAB.416747},
	number = {4},
	urldate = {2024-01-22},
	journal = {JOSA B},
	author = {Wolff, C. and Smith, M. J. A. and Stiller, B. and Poulton, C. G.},
	month = apr,
	year = {2021},
	note = {Publisher: Optica Publishing Group},
	pages = {1243--1269},
}

@book{BoydBook,
	address = {New York},
	author = {Boyd, RW},
	booktitle = {Nonlinear Optics--3rd Edition},
	year = {2007},
    publisher={Springer}
}

@article{Kozii2017,
  title = {Non-Hermitian topological theory of finite-lifetime quasiparticles: Prediction of bulk Fermi arc due to exceptional point},
  author = {Kozii, Vladyslav and Fu, Liang},
  journal = {Phys. Rev. B},
  volume = {109},
  issue = {23},
  pages = {235139},
  numpages = {5},
  year = {2024},
  month = {Jun},
  publisher = {American Physical Society},
  doi = {10.1103/PhysRevB.109.235139},
  url = {https://link.aps.org/doi/10.1103/PhysRevB.109.235139}
}

@article{Zhou2018,
  title={Observation of bulk Fermi arc and polarization half charge from paired exceptional points},
  author={Zhou, Hengyun and Peng, Chao and Yoon, Yoseob and Hsu, Chia Wei and Nelson, Keith A and Fu, Liang and Joannopoulos, John D and Solja{\v{c}}i{\'c}, Marin and Zhen, Bo},
  journal={Science},
  volume={359},
  number={6379},
  pages={1009--1012},
  year={2018},
  publisher={American Association for the Advancement of Science},
  url = {https://doi.org/10.1126/science.aap9859}
}

@article{El-Ganainy2018,
	author = {El-Ganainy, Ramy and Makris, Konstantinos G. and Khajavikhan, Mercedeh and Musslimani, Ziad H. and Rotter, Stefan and Christodoulides, Demetrios N.},
	date = {2018/01/01},
	doi = {10.1038/nphys4323},
	id = {El-Ganainy2018},
	isbn = {1745-2481},
	journal = {Nature Physics},
	number = {1},
	pages = {11--19},
	title = {Non-Hermitian physics and PT symmetry},
	url = {https://doi.org/10.1038/nphys4323},
	volume = {14},
	year = {2018}}

@article{Bergholtz2021,
  title = {Exceptional topology of non-Hermitian systems},
  author = {Bergholtz, Emil J. and Budich, Jan Carl and Kunst, Flore K.},
  journal = {Rev. Mod. Phys.},
  volume = {93},
  issue = {1},
  pages = {015005},
  numpages = {31},
  year = {2021},
  month = {Feb},
  publisher = {American Physical Society},
  doi = {10.1103/RevModPhys.93.015005},
  url = {https://link.aps.org/doi/10.1103/RevModPhys.93.015005}
}

@book{KatoBook,
	address = {Berlin},
	author = {Kato, T.},
	booktitle = {Perturbation theory of linear operators},
	editor = {Cappelli, Andrea and Mussardo, Giuseppe},
	publisher = {Springer},
	title = {{Perturbation theory of linear operators}},
	year = {1966}}

@article{Heiss2012,
	author = {W D Heiss},
	doi = {10.1088/1751-8113/45/44/444016},
	journal = {Journal of Physics A: Mathematical and Theoretical},
	month = {oct},
	number = {44},
	pages = {444016},
	publisher = {IOP Publishing},
	title = {The physics of exceptional points},
	url = {https://dx.doi.org/10.1088/1751-8113/45/44/444016},
	volume = {45},
	year = {2012}}

@article{Miri2019,
	author = {Mohammad-Ali Miri and Andrea Al{\`u}},
	doi = {10.1126/science.aar7709},
	journal = {Science},
	number = {6422},
	pages = {eaar7709},
	title = {Exceptional points in optics and photonics},
	url = {https://www.science.org/doi/abs/10.1126/science.aar7709},
	volume = {363},
	year = {2019}}

@article{Ashida2020,
	author = {Yuto Ashida and Zongping Gong and Masahito Ueda},
	doi = {10.1080/00018732.2021.1876991},
	journal = {Advances in Physics},
	number = {3},
	pages = {249-435},
	publisher = {Taylor & Francis},
	title = {Non-Hermitian physics},
	url = {https://doi.org/10.1080/00018732.2021.1876991},
	volume = {69},
	year = {2020}}

@article{Berry2004,
	author = {Berry, M. V.},
	doi = {10.1023/B:CJOP.0000044002.05657.04},
	issn = {1572-9486},
	journal = {Czech. J. Phys.},
	month = {oct},
	number = {10},
	pages = {1039--1047},
	title = {{Physics of Nonhermitian Degeneracies}},
	url = {https://doi.org/10.1023/B:CJOP.0000044002.05657.04},
	volume = {54},
	year = {2004}}

@article{Carlstrom2018,
  title = {Exceptional links and twisted Fermi ribbons in non-Hermitian systems},
  author = {Carlstr\"om, Johan and Bergholtz, Emil J.},
  journal = {Phys. Rev. A},
  volume = {98},
  issue = {4},
  pages = {042114},
  numpages = {5},
  year = {2018},
  month = {Oct},
  publisher = {American Physical Society},
  doi = {10.1103/PhysRevA.98.042114},
  url = {https://link.aps.org/doi/10.1103/PhysRevA.98.042114}
}

@article{Lin2011,
  title = {Unidirectional Invisibility Induced by $\mathcal{P}\mathcal{T}$-Symmetric Periodic Structures},
  author = {Lin, Zin and Ramezani, Hamidreza and Eichelkraut, Toni and Kottos, Tsampikos and Cao, Hui and Christodoulides, Demetrios N.},
  journal = {Phys. Rev. Lett.},
  volume = {106},
  issue = {21},
  pages = {213901},
  numpages = {4},
  year = {2011},
  month = {May},
  publisher = {American Physical Society},
  doi = {10.1103/PhysRevLett.106.213901},
  url = {https://link.aps.org/doi/10.1103/PhysRevLett.106.213901}
}

@article{Brandstetter2014,
  title={Reversing the pump dependence of a laser at an exceptional point},
  author={Brandstetter, M and Liertzer, M and Deutsch, C and Klang, P and Sch{\"o}berl, J and T{\"u}reci, Hakan E and Strasser, G and Unterrainer, K and Rotter, S},
  journal={Nature communications},
  volume={5},
  number={1},
  pages={4034},
  year={2014},
  publisher={Nature Publishing Group UK London},
  doi={https://doi.org/10.1038/ncomms5034}
}

@ARTICLE{Peng2016,
       author = {{Peng}, Bo and {{\"O}zdemir}, {\c{S}}ahin Kaya and {Liertzer}, Matthias and {Chen}, Weijian and {Kramer}, Johannes and {Y{\i}lmaz}, Huzeyfe and {Wiersig}, Jan and {Rotter}, Stefan and {Yang}, Lan},
        title = "{Chiral modes and directional lasing at exceptional points}",
      journal = {Proceedings of the National Academy of Science},
         year = 2016,
        month = jun,
       volume = {113},
       number = {25},
        pages = {6845-6850},
          doi = {10.1073/pnas.1603318113},
       adsurl = {https://ui.adsabs.harvard.edu/abs/2016PNAS..113.6845P}
}

@article{Goldzak2018,
  title = {Light Stops at Exceptional Points},
  author = {Goldzak, Tamar and Mailybaev, Alexei A. and Moiseyev, Nimrod},
  journal = {Phys. Rev. Lett.},
  volume = {120},
  issue = {1},
  pages = {013901},
  numpages = {5},
  year = {2018},
  month = {Jan},
  publisher = {American Physical Society},
  doi = {10.1103/PhysRevLett.120.013901},
  url = {https://link.aps.org/doi/10.1103/PhysRevLett.120.013901}
}

@article{Feng2013,
  title={Demonstration of a large-scale optical exceptional point structure},
  author={Feng, Liang and Zhu, Xuefeng and Yang, Sui and Zhu, Hanyu and Zhang, Peng and Yin, Xiaobo and Wang, Yuan and Zhang, Xiang},
  journal={Optics express},
  volume={22},
  number={2},
  pages={1760--1767},
  year={2013},
  publisher={Optical Society of America},
  doi={https://doi.org/10.1364/OE.22.001760}
}

@article{Feng2013a,
  title={Experimental demonstration of a unidirectional reflectionless parity-time metamaterial at optical frequencies},
  author={Feng, Liang and Xu, Ye-Long and Fegadolli, William S and Lu, Ming-Hui and Oliveira, Jos{\'e} EB and Almeida, Vilson R and Chen, Yan-Feng and Scherer, Axel},
  journal={Nature materials},
  volume={12},
  number={2},
  pages={108--113},
  year={2013},
  publisher={Nature Publishing Group UK London},
  doi={https://doi.org/10.1038/nmat3495}
}

@article{Zhen2015,
  title={Spawning rings of exceptional points out of Dirac cones},
  author={Zhen, Bo and Hsu, Chia Wei and Igarashi, Yuichi and Lu, Ling and Kaminer, Ido and Pick, Adi and Chua, Song-Liang and Joannopoulos, John D and Solja{\v{c}}i{\'c}, Marin},
  journal={Nature},
  volume={525},
  number={7569},
  pages={354--358},
  year={2015},
  publisher={Nature Publishing Group UK London},
  doi={https://doi.org/10.1038/nature14889}
}

@article{Guo2009,
  title = {Observation of $\mathcal{P}\mathcal{T}$-Symmetry Breaking in Complex Optical Potentials},
  author = {Guo, A. and Salamo, G. J. and Duchesne, D. and Morandotti, R. and Volatier-Ravat, M. and Aimez, V. and Siviloglou, G. A. and Christodoulides, D. N.},
  journal = {Phys. Rev. Lett.},
  volume = {103},
  issue = {9},
  pages = {093902},
  numpages = {4},
  year = {2009},
  month = {Aug},
  publisher = {American Physical Society},
  doi = {10.1103/PhysRevLett.103.093902},
  url = {https://link.aps.org/doi/10.1103/PhysRevLett.103.093902}
}

@article{Ruter2010,
  title={Observation of parity--time symmetry in optics},
  author={R{\"u}ter, Christian E and Makris, Konstantinos G and El-Ganainy, Ramy and Christodoulides, Demetrios N and Segev, Mordechai and Kip, Detlef},
  journal={Nature physics},
  volume={6},
  number={3},
  pages={192--195},
  year={2010},
  publisher={Nature Publishing Group UK London},
  doi={https://doi.org/10.1038/nphys1515}
}

@article{ozdemir_paritytime_2019,
  title = {Parity--Time Symmetry and Exceptional Points in Photonics},
  author = {{\"O}zdemir, {\c S} K. and Rotter, S. and Nori, F. and Yang, L.},
  year = {2019},
  month = aug,
  journal = {Nature Materials},
  volume = {18},
  number = {8},
  pages = {783--798},
  publisher = {Nature Publishing Group},
  issn = {1476-4660},
  doi = {10.1038/s41563-019-0304-9},
  copyright = {2019 The Author(s), under exclusive licence to Springer Nature Limited},
  langid = {english}
}

@article{niklesSimpleDistributedFiber1996,
  title = {Simple Distributed Fiber Sensor Based on {{Brillouin}} Gain Spectrum Analysis},
  author = {Nikl{\`e}s, Marc and Th{\'e}venaz, Luc and Robert, Philippe A.},
  year = {1996},
  month = may,
  journal = {Optics Letters},
  volume = {21},
  number = {10},
  pages = {758--760},
  publisher = {Optica Publishing Group},
  issn = {1539-4794},
  doi = {10.1364/OL.21.000758},
  urldate = {2023-03-22},
  copyright = {{\copyright} 1996 Optical Society of America},
  langid = {english}
}

@inproceedings{hasegawaMeasurementBrillouinGain1999,
  title = {Measurement of {{Brillouin}} Gain Spectrum Distribution along an Optical Fiber by Direct Frequency Modulation of a Laser Diode},
  booktitle = {Fiber {{Optic Sensor Technology}} and {{Applications}}},
  author = {Hasegawa, Takemi and Hotate, Kazuo},
  year = {1999},
  month = dec,
  volume = {3860},
  pages = {306--316},
  publisher = {SPIE},
  doi = {10.1117/12.372976},
  urldate = {2025-05-15}
}

@article{geilenExtremeThermodynamicsNanolitre2023,
  title = {Extreme Thermodynamics in Nanolitre Volumes through Stimulated {{Brillouin}}--{{Mandelstam}} Scattering},
  author = {Geilen, Andreas and Popp, Alexandra and Das, Debayan and Junaid, Saher and Poulton, Christopher G. and Chemnitz, Mario and Marquardt, Christoph and Schmidt, Markus A. and Stiller, Birgit},
  year = {2023},
  month = sep,
  journal = {Nature Physics},
  pages = {1--8},
  publisher = {Nature Publishing Group},
  issn = {1745-2481},
  doi = {10.1038/s41567-023-02205-1},
  urldate = {2023-12-04},
  copyright = {2023 The Author(s)},
  langid = {english}
}

@article{marpaungIntegratedMicrowavePhotonics2019,
  title = {Integrated Microwave Photonics},
  author = {Marpaung, David and Yao, Jianping and Capmany, Jos{\'e}},
  year = {2019},
  month = feb,
  journal = {Nature Photonics},
  volume = {13},
  number = {2},
  pages = {80--90},
  publisher = {Nature Publishing Group},
  issn = {1749-4893},
  doi = {10.1038/s41566-018-0310-5},
  urldate = {2023-05-30},
  copyright = {2019 Springer Nature Limited},
  langid = {english}
}

@article{Kabakova2024,
  title={Brillouin microscopy},
  author={Kabakova, Irina and Zhang, Jitao and Xiang, Yuchen and Caponi, Silvia and Bilenca, Alberto and Guck, Jochen and Scarcelli, Giuliano},
  journal={Nature Reviews Methods Primers},
  volume={4},
  number={1},
  pages={8},
  year={2024},
  doi = {10.1038/s43586-023-00286-z},
  publisher={Nature Publishing Group UK London}
}

@misc{Prevedel2019, author = {Robert Prevedel and Alba Diz-Muñoz and Giancarlo Ruocco and Giuseppe Antonacci}, doi = {10.1038/s41592-019-0543-3}, issn = {15487105}, issue = {10}, journal = {Nature Methods}, title = {Brillouin microscopy: an emerging tool for mechanobiology}, volume = {16}, year = {2019} }

@article{otterstromSiliconBrillouinLaser2018a,
  title = {A Silicon {{Brillouin}} Laser},
  author = {Otterstrom, Nils T. and Behunin, Ryan O. and Kittlaus, Eric A. and Wang, Zheng and Rakich, Peter T.},
  year = {2018},
  month = jun,
  journal = {Science},
  volume = {360},
  number = {6393},
  pages = {1113--1116},
  publisher = {American Association for the Advancement of Science},
  doi = {10.1126/science.aar6113},
  urldate = {2025-05-15}
}

@article{Gundavarapu2019,
  title={Sub-hertz fundamental linewidth photonic integrated Brillouin laser},
  author={Gundavarapu, Sarat and Brodnik, Grant M and Puckett, Matthew and Huffman, Taran and Bose, Debapam and Behunin, Ryan and Wu, Jianfeng and Qiu, Tiequn and Pinho, C{\'a}tia and Chauhan, Nitesh and others},
  journal={Nature Photonics},
  volume={13},
  number={1},
  pages={60--67},
  year={2019},
  publisher={Nature Publishing Group UK London},
  doi={https://doi.org/10.1038/s41566-018-0313-2}
}

@article{zengOpticalVortexBrillouin2023,
  title = {Optical {{Vortex Brillouin Laser}}},
  author = {Zeng, Xinglin and Russell, Philip St. J. and Chen, Yang and Wang, Zheqi and Wong, Gordon K. L. and Roth, Paul and Frosz, Michael H. and Stiller, Birgit},
  year = {2023},
  journal = {Laser \& Photonics Reviews},
  volume = {17},
  number = {4},
  pages = {2200277},
  issn = {1863-8899},
  doi = {10.1002/lpor.202200277},
  urldate = {2023-05-30},
  copyright = {{\copyright} 2023 The Authors. Laser \& Photonics Reviews published by Wiley-VCH GmbH},
  langid = {english}
}

@article{zhuStoredLightOptical2007,
  title = {Stored {{Light}} in an {{Optical Fiber}} via {{Stimulated Brillouin Scattering}}},
  author = {Zhu, Zhaoming and Gauthier, Daniel J. and Boyd, Robert W.},
  year = {2007},
  month = dec,
  journal = {Science},
  volume = {318},
  number = {5857},
  pages = {1748--1750},
  publisher = {American Association for the Advancement of Science},
  doi = {10.1126/science.1149066},
  urldate = {2023-02-02}
}

@article{geilenHighSpeedCoherentPhotonic2024,
  title = {High-{{Speed Coherent Photonic Random-Access Memory}} in {{Long-Lasting Sound Waves}}},
  author = {Geilen, Andreas and Becker, Steven and Stiller, Birgit},
  year = {2024},
  month = nov,
  journal = {ACS Photonics},
  volume = {11},
  number = {11},
  pages = {4524--4532},
  publisher = {American Chemical Society},
  doi = {10.1021/acsphotonics.4c00478},
  urldate = {2025-05-15}
}

@article{merkleinChipintegratedCoherentPhotonicphononic2017a,
  title = {A Chip-Integrated Coherent Photonic-Phononic Memory},
  author = {Merklein, Moritz and Stiller, Birgit and Vu, Khu and Madden, Stephen J. and Eggleton, Benjamin J.},
  year = {2017},
  month = sep,
  journal = {Nature Communications},
  volume = {8},
  number = {1},
  pages = {574},
  publisher = {Nature Publishing Group},
  issn = {2041-1723},
  doi = {10.1038/s41467-017-00717-y},
  urldate = {2025-05-15},
  copyright = {2017 The Author(s)},
  langid = {english}
}

@article{safferBrillouinbasedStorageQPSK2024,
  title={Brillouin-based storage of QPSK signals with fully tunable phase retrieval},
  author={Saffer, Olivia and Marines Cabello, Jes{\'u}s Humberto and Becker, Steven and Geilen, Andreas and Stiller, Birgit},
  journal={APL Photonics},
  volume={10},
  number={6},
  pages = {060802},
  year={2025},
  publisher={AIP Publishing},
  doi={doi.org/10.1063/5.0241508}
}

@article{stillerCoherentlyRefreshingHypersonic2020,
  title = {Coherently Refreshing Hypersonic Phonons for Light Storage},
  author = {Stiller, Birgit and Merklein, Moritz and Wolff, Christian and Vu, Khu and Ma, Pan and Madden, Stephen J. and Eggleton, Benjamin J.},
  year = {2020},
  month = may,
  journal = {Optica},
  volume = {7},
  number = {5},
  pages = {492--497},
  publisher = {Optica Publishing Group},
  issn = {2334-2536},
  doi = {10.1364/OPTICA.386535},
  urldate = {2023-05-30},
  copyright = {{\copyright} 2020 Optical Society of America},
  langid = {english}
}

@article{beckerOptoacousticFieldprogrammablePerceptron2024,
  title = {An Optoacoustic Field-Programmable Perceptron for Recurrent Neural Networks},
  author = {Becker, Steven and Englund, Dirk and Stiller, Birgit},
  year = {2024},
  month = apr,
  journal = {Nature Communications},
  volume = {15},
  number = {1},
  pages = {3020},
  publisher = {Nature Publishing Group},
  issn = {2041-1723},
  doi = {10.1038/s41467-024-47053-6},
  urldate = {2025-05-15},
  copyright = {2024 The Author(s)},
  langid = {english}
}

@article{slinkovAllopticalNonlinearActivation2025,
url = {https://doi.org/10.1515/nanoph-2024-0513},
title = {All-optical nonlinear activation function based on stimulated Brillouin scattering},
author = {Grigorii Slinkov and Steven Becker and Dirk Englund and Birgit Stiller},
pages = {2711--2722},
volume = {14},
number = {16},
journal = {Nanophotonics},
doi = {doi:10.1515/nanoph-2024-0513},
year = {2025}
}

@article{stillerPhotonicCrystalFiber2010,
  title = {Photonic Crystal Fiber Mapping Using {{Brillouin}} Echoes Distributed Sensing},
  author = {Stiller, B. and Foaleng, S. M. and Beugnot, J.-C. and Lee, M. W. and Delqu{\'e}, M. and Bouwmans, G. and Kudlinski, A. and Th{\'e}venaz, L. and Maillotte, H. and Sylvestre, T.},
  year = {2010},
  month = sep,
  journal = {Optics Express},
  volume = {18},
  number = {19},
  pages = {20136--20142},
  publisher = {Optica Publishing Group},
  issn = {1094-4087},
  doi = {10.1364/OE.18.020136},
  urldate = {2025-05-28},
  copyright = {{\copyright} 2010 OSA},
  langid = {english}
}

@article{godetBrillouinSpectroscopyOptical2017,
  title = {Brillouin Spectroscopy of Optical Microfibers and Nanofibers},
  author = {Godet, Adrien and Ndao, Abdoulaye and Sylvestre, Thibaut and Pecheur, Vincent and Lebrun, Sylvie and Pauliat, Gilles and Beugnot, Jean-Charles and Huy, Kien Phan},
  year = {2017},
  month = oct,
  journal = {Optica},
  volume = {4},
  number = {10},
  pages = {1232--1238},
  publisher = {Optica Publishing Group},
  issn = {2334-2536},
  doi = {10.1364/OPTICA.4.001232},
  urldate = {2025-05-16},
  copyright = {{\copyright} 2017 Optical Society of America},
  langid = {english}
}

@article{eggletonBrillouinIntegratedPhotonics2019,
  title = {Brillouin Integrated Photonics},
  author = {Eggleton, Benjamin J. and Poulton, Christopher G. and Rakich, Peter T. and Steel, Michael J. and Bahl, Gaurav},
  year = {2019},
  month = oct,
  journal = {Nature Photonics},
  volume = {13},
  number = {10},
  pages = {664--677},
  publisher = {Nature Publishing Group},
  issn = {1749-4893},
  doi = {10.1038/s41566-019-0498-z},
  urldate = {2025-05-16},
  copyright = {2019 Springer Nature Limited},
  langid = {english}
}

@article{kittlausLargeBrillouinAmplification2016,
  title = {Large {{Brillouin}} Amplification in Silicon},
  author = {Kittlaus, Eric A. and Shin, Heedeuk and Rakich, Peter T.},
  year = {2016},
  month = jul,
  journal = {Nature Photonics},
  volume = {10},
  number = {7},
  pages = {463--467},
  publisher = {Nature Publishing Group},
  issn = {1749-4893},
  doi = {10.1038/nphoton.2016.112},
  urldate = {2025-05-16},
  copyright = {2016 Springer Nature Limited},
  langid = {english}
}

@article{kobyakovStimulatedBrillouinScattering2010a,
  title = {Stimulated {{Brillouin}} Scattering in Optical Fibers},
  author = {Kobyakov, Andrey and Sauer, Michael and Chowdhury, Dipak},
  year = {2010},
  month = mar,
  journal = {Advances in Optics and Photonics},
  volume = {2},
  number = {1},
  pages = {1--59},
  publisher = {Optica Publishing Group},
  issn = {1943-8206},
  doi = {10.1364/AOP.2.000001},
  urldate = {2025-07-01},
  copyright = {{\copyright} 2009 Optical Society of America},
  langid = {english}
}

@article{neijtsOnchipStimulatedBrillouin2024,
  title = {On-Chip Stimulated {{Brillouin}} Scattering via Surface Acoustic Waves},
  author = {Neijts, Govert and Lai, Choon Kong and Riseng, Maren Kramer and Choi, Duk-Yong and Yan, Kunlun and Marpaung, David and Madden, Stephen J. and Eggleton, Benjamin J. and Merklein, Moritz},
  year = {2024},
  month = oct,
  journal = {APL Photonics},
  volume = {9},
  number = {10},
  pages = {106114},
  issn = {2378-0967},
  doi = {10.1063/5.0220496},
  urldate = {2025-05-15}
}

@article{rodriguesCrossPolarizedStimulatedBrillouin2025,
  title = {Cross-{{Polarized Stimulated Brillouin Scattering}} in {{Lithium Niobate Waveguides}}},
  author = {Rodrigues, Caique C. and Schilder, Nick J. and Zurita, Roberto O. and Magalh{\~a}es, Let{\'i}cia S. and {Shams-Ansari}, Amirhassan and {dos Santos}, Felipe J. L. and Paiano, Ot{\'a}vio M. and Alegre, Thiago P. M. and Lon{\v c}ar, Marko and Wiederhecker, Gustavo S.},
  year = {2025},
  month = mar,
  journal = {Physical Review Letters},
  volume = {134},
  number = {11},
  pages = {113601},
  publisher = {American Physical Society},
  doi = {10.1103/PhysRevLett.134.113601},
  urldate = {2025-05-15}
}

@article{
Ye2025,
author = {Kaixuan Ye  and Hanke Feng  and Randy te Morsche  and Chuangchuang Wei  and Yvan Klaver  and Akhileshwar Mishra  and Zheng Zheng  and Akshay Keloth  and Ahmet Tarık Işık  and Zhaoxi Chen  and Cheng Wang  and David Marpaung },
title = {Integrated Brillouin photonics in thin-film lithium niobate},
journal = {Science Advances},
volume = {11},
number = {18},
pages = {eadv4022},
year = {2025},
doi = {10.1126/sciadv.adv4022}}

@article{xuStrongOptomechanicalInteractions2023,
  title = {Strong Optomechanical Interactions with Long-Lived Fundamental Acoustic Waves},
  author = {Xu, Wendao and Iyer, Arjun and Jin, Lei and Set, Sze Y. and Renninger, William H.},
  year = {2023},
  month = feb,
  journal = {Optica},
  volume = {10},
  number = {2},
  pages = {206--213},
  publisher = {Optica Publishing Group},
  issn = {2334-2536},
  doi = {10.1364/OPTICA.476764},
  urldate = {2025-05-14},
  copyright = {{\copyright} 2023 Optica Publishing Group},
  langid = {english}
}

@article{renningerBulkCrystallineOptomechanics2018a,
  title = {Bulk Crystalline Optomechanics},
  author = {Renninger, W. H. and Kharel, P. and Behunin, R. O. and Rakich, P. T.},
  year = {2018},
  month = jun,
  journal = {Nature Physics},
  volume = {14},
  number = {6},
  pages = {601--607},
  publisher = {Nature Publishing Group},
  issn = {1745-2481},
  doi = {10.1038/s41567-018-0090-3},
  urldate = {2025-07-01},
  copyright = {2018 The Author(s)},
  langid = {english}
}

@article{
Chu2017,
author = {Yiwen Chu  and Prashanta Kharel  and William H. Renninger  and Luke D. Burkhart  and Luigi Frunzio  and Peter T. Rakich  and Robert J. Schoelkopf },
title = {Quantum acoustics with superconducting qubits},
journal = {Science},
volume = {358},
number = {6360},
pages = {199-202},
year = {2017},
doi = {10.1126/science.aao1511}
}

@article{PhysRevResearch.5.043140,
  title = {Brillouin optomechanics in the quantum ground state},
  author = {Doeleman, H. M. and Schatteburg, T. and Benevides, R. and Vollenweider, S. and Macri, D. and Chu, Y.},
  journal = {Phys. Rev. Res.},
  volume = {5},
  issue = {4},
  pages = {043140},
  numpages = {28},
  year = {2023},
  month = {Nov},
  publisher = {American Physical Society},
  doi = {10.1103/PhysRevResearch.5.043140},
  url = {https://link.aps.org/doi/10.1103/PhysRevResearch.5.043140}
}

@article{blazquezmartinezOptoacousticCoolingTraveling2024a,
  title = {Optoacoustic {{Cooling}} of {{Traveling Hypersound Waves}}},
  author = {Bl{\'a}zquez Mart{\'i}nez, Laura and Wiedemann, Philipp and Zhu, Changlong and Geilen, Andreas and Stiller, Birgit},
  year = {2024},
  month = jan,
  journal = {Physical Review Letters},
  volume = {132},
  number = {2},
  pages = {023603},
  publisher = {American Physical Society},
  doi = {10.1103/PhysRevLett.132.023603},
  urldate = {2025-03-27}
}

@article{cryer-jenkinsBrillouinMandelstamScattering2025b,
  title = {Brillouin--{{Mandelstam}} Scattering in Telecommunications Optical Fiber at Millikelvin Temperatures},
  author = {{Cryer-Jenkins}, E. A. and Leung, A. C. and Rathee, H. and Tan, A. K. C. and Major, K. D. and Vanner, M. R.},
  year = {2025},
  month = jan,
  journal = {APL Photonics},
  volume = {10},
  number = {1},
  pages = {010805},
  issn = {2378-0967},
  doi = {10.1063/5.0241253},
  urldate = {2025-05-15}
}

@article{enzianObservationBrillouinOptomechanical2019b,
  title = {Observation of {{Brillouin}} Optomechanical Strong Coupling with an 11 {{GHz}} Mechanical Mode},
  author = {Enzian, G. and Szczykulska, M. and Silver, J. and Bino, L. Del and Zhang, S. and Walmsley, I. A. and Del'Haye, P. and Vanner, M. R.},
  year = {2019},
  month = jan,
  journal = {Optica},
  volume = {6},
  number = {1},
  pages = {7--14},
  publisher = {Optica Publishing Group},
  issn = {2334-2536},
  doi = {10.1364/OPTICA.6.000007},
  urldate = {2024-07-12},
  langid = {english}
}

@article{PhysRevLett.126.033601,
  title = {Single-Phonon Addition and Subtraction to a Mechanical Thermal State},
  author = {Enzian, G. and Price, J. J. and Freisem, L. and Nunn, J. and Janousek, J. and Buchler, B. C. and Lam, P. K. and Vanner, M. R.},
  journal = {Phys. Rev. Lett.},
  volume = {126},
  issue = {3},
  pages = {033601},
  numpages = {6},
  year = {2021},
  month = {Jan},
  publisher = {American Physical Society},
  doi = {10.1103/PhysRevLett.126.033601},
  url = {https://link.aps.org/doi/10.1103/PhysRevLett.126.033601}
}

@article{zhangQuantumCoherentControl2023b,
  title = {Quantum Coherent Control in Pulsed Waveguide Optomechanics},
  author = {Zhang, Junyin and Zhu, Changlong and Wolff, Christian and Stiller, Birgit},
  year = {2023},
  month = jan,
  journal = {Physical Review Research},
  volume = {5},
  number = {1},
  pages = {013010},
  publisher = {American Physical Society},
  doi = {10.1103/PhysRevResearch.5.013010},
  urldate = {2025-03-06}
}

@article{zhuOptoacousticEntanglementContinuous2024,
  title = {Optoacoustic {{Entanglement}} in a {{Continuous Brillouin-Active Solid State System}}},
  author = {Zhu, Changlong and Genes, Claudiu and Stiller, Birgit},
  year = {2024},
  month = nov,
  journal = {Physical Review Letters},
  volume = {133},
  number = {20},
  pages = {203602},
  publisher = {American Physical Society},
  doi = {10.1103/PhysRevLett.133.203602},
  urldate = {2025-02-14}
}

@article{StrongCoupling2025,
      title={Cavity-less Brillouin strong coupling in a solid-state continuous system}, 
      author={Laura Blázquez Martínez and Changlong Zhu and Birgit Stiller},
      year={2025},
      journal={arXiv:2507.08673},
      url={https://arxiv.org/abs/2507.08673}, 
}

@article{PhysRevResearch.5.L042010,
  title = {Braid-protected topological band structures with unpaired exceptional points},
  author = {K\"onig, J. Lukas K. and Yang, Kang and Budich, Jan Carl and Bergholtz, Emil J.},
  journal = {Phys. Rev. Res.},
  volume = {5},
  issue = {4},
  pages = {L042010},
  numpages = {6},
  year = {2023},
  month = {Oct},
  publisher = {American Physical Society},
  doi = {10.1103/PhysRevResearch.5.L042010},
  url = {https://link.aps.org/doi/10.1103/PhysRevResearch.5.L042010}
}

@article{wang2021topological,
  title={Topological complex-energy braiding of non-Hermitian bands},
  author={Wang, Kai and Dutt, Avik and Wojcik, Charles C and Fan, Shanhui},
  journal={Nature},
  volume={598},
  number={7879},
  pages={59--64},
  year={2021},
  publisher={Nature Publishing Group UK London},
  doi={https://doi.org/10.1038/s41586-021-03848-x}
}

@article{Ippen1972,
    author = {Ippen, E.P. and Stolen, R.H.},
    title = {Stimulated Brillouin scattering in optical fibers},
    journal = {Applied Physics Letters},
    volume = {21},
    number = {11},
    pages = {539-541},
    year = {1972},
    month = {12},
    issn = {0003-6951},
    doi = {10.1063/1.1654249},
    url = {https://doi.org/10.1063/1.1654249}
}

@article{chen_quantum_2025,
  title = {Quantum Tomography of a Third-Order Exceptional Point in a Dissipative Trapped Ion},
  author = {Chen, Y.-Y. and Li, K. and Zhang, L. and Wu, Y.-K. and Ma, J.-Y. and Yang, H.-X. and Zhang, C. and Qi, B.-X. and Zhou, Z.-C. and Hou, P.-Y. and Xu, Y. and Duan, L.-M.},
  year = {2025},
  month = aug,
  journal = {Nature Communications},
  volume = {16},
  number = {1},
  pages = {7478},
  publisher = {Nature Publishing Group},
  issn = {2041-1723},
  doi = {10.1038/s41467-025-62573-5},
  copyright = {2025 The Author(s)},
  langid = {english}
}

@article{han_measuring_2024,
  title = {Measuring Topological Invariants for Higher-Order Exceptional Points in Quantum Three-Mode Systems},
  author = {Han, Pei-Rong and Ning, Wen and Huang, Xin-Jie and Zheng, Ri-Hua and Yang, Shou-Bang and Wu, Fan and Yang, Zhen-Biao and Su, Qi-Ping and Yang, Chui-Ping and Zheng, Shi-Biao},
  year = {2024},
  month = nov,
  journal = {Nature Communications},
  volume = {15},
  number = {1},
  pages = {10293},
  publisher = {Nature Publishing Group},
  issn = {2041-1723},
  doi = {10.1038/s41467-024-54662-8},
  copyright = {2024 The Author(s)},
  langid = {english}
}

@article{hodaei_enhanced_2017,
  title = {Enhanced Sensitivity at Higher-Order Exceptional Points},
  author = {Hodaei, Hossein and Hassan, Absar U. and Wittek, Steffen and {Garcia-Gracia}, Hipolito and {El-Ganainy}, Ramy and Christodoulides, Demetrios N. and Khajavikhan, Mercedeh},
  year = {2017},
  month = aug,
  journal = {Nature},
  volume = {548},
  number = {7666},
  pages = {187--191},
  publisher = {Nature Publishing Group},
  issn = {1476-4687},
  doi = {10.1038/nature23280},
  copyright = {2017 Macmillan Publishers Limited, part of Springer Nature. All rights reserved.},
  langid = {english}
}

@article{jahangiri_observation_2025,
  title = {Observation of Anti-{{PT-symmetry}} and Higher-Order Exceptional Point {{PT-symmetry}} in Ternary Systems for Single-Mode Operation},
  author = {Jahangiri, Milad and Parsanasab, Gholam-Mohammad and Hajshahvaladi, Leila},
  year = {2025},
  month = feb,
  journal = {Scientific Reports},
  volume = {15},
  number = {1},
  pages = {4823},
  publisher = {Nature Publishing Group},
  issn = {2045-2322},
  doi = {10.1038/s41598-025-85623-w},
  copyright = {2025 The Author(s)},
  langid = {english}
}

@article{li_programmable_2024,
  title={Programmable simulation of high-order exceptional point with a trapped ion},
  author={Li, Yue and Wu, Yang and Zhou, Yuqi and Zhang, Mengxiang and Zhao, Xingyu and Yuan, Yibo and Cheng, Xu and Li, Yi and Qin, Xi and Rong, Xing and others},
  journal={Quantum Frontiers},
  volume={4},
  number={1},
  pages={16},
  year={2025},
  publisher={Springer},
  doi={https://doi.org/10.1007/s44214-025-00088-2}
}

@article{zhang_experimental_2025,
  title = {Experimental Observation of Non-Markovian Quantum Exceptional Points},
  author = {Zhang, Hao-Long and Han, Pei-Rong and Wu, Fan and Ning, Wen and Yang, Zhen-Biao and Zheng, Shi-Biao},
  journal = {Phys. Rev. Lett.},
  volume = {135},
  issue = {23},
  pages = {230203},
  numpages = {7},
  year = {2025},
  month = {Dec},
  publisher = {American Physical Society},
  doi = {10.1103/jk6y-55xp},
  url = {https://link.aps.org/doi/10.1103/jk6y-55xp}
}

@article{tang_exceptional_2020,
  title = {Exceptional Nexus with a Hybrid Topological Invariant},
  author = {Tang, Weiyuan and Jiang, Xue and Ding, Kun and Xiao, Yi-Xin and Zhang, Zhao-Qing and Chan, C. T. and Ma, Guancong},
  year = {2020},
  month = nov,
  journal = {Science},
  volume = {370},
  number = {6520},
  pages = {1077--1080},
  publisher = {American Association for the Advancement of Science},
  doi = {10.1126/science.abd8872}
}

@article{tang_realization_2023,
  title = {Realization and Topological Properties of Third-Order Exceptional Lines Embedded in Exceptional Surfaces},
  author = {Tang, Weiyuan and Ding, Kun and Ma, Guancong},
  year = {2023},
  month = oct,
  journal = {Nature Communications},
  volume = {14},
  number = {1},
  pages = {6660},
  publisher = {Nature Publishing Group},
  issn = {2041-1723},
  doi = {10.1038/s41467-023-42414-z},
  copyright = {2023 The Author(s)},
  langid = {english}
}

@article{wang_exceptional_2024,
  title = {Exceptional {{Nexus}} in {{Bose-Einstein Condensates}} with {{Collective Dissipation}}},
  author = {Wang, Chenhao and Li, Nan and Xie, Jin and Ding, Cong and Ji, Zhonghua and Xiao, Liantuan and Jia, Suotang and Yan, Bo and Hu, Ying and Zhao, Yanting},
  year = {2024},
  month = jun,
  journal = {Physical Review Letters},
  volume = {132},
  number = {25},
  pages = {253401},
  publisher = {American Physical Society},
  doi = {10.1103/PhysRevLett.132.253401}
}

@article{wang_experimental_2023,
  title = {Experimental Simulation of Symmetry-Protected Higher-Order Exceptional Points with Single Photons},
  author = {Wang, Kunkun and Xiao, Lei and Lin, Haiqing and Yi, Wei and Bergholtz, Emil J. and Xue, Peng},
  year = {2023},
  month = aug,
  journal = {Science Advances},
  volume = {9},
  number = {34},
  pages = {eadi0732},
  publisher = {American Association for the Advancement of Science},
  doi = {10.1126/sciadv.adi0732}
}

@article{wu_third-order_2024,
  title = {Third-Order Exceptional Line in a Nitrogen-Vacancy Spin System},
  author = {Wu, Yang and Wang, Yunhan and Ye, Xiangyu and Liu, Wenquan and Niu, Zhibo and Duan, Chang-Kui and Wang, Ya and Rong, Xing and Du, Jiangfeng},
  year = {2024},
  month = feb,
  journal = {Nature Nanotechnology},
  volume = {19},
  number = {2},
  pages = {160--165},
  publisher = {Nature Publishing Group},
  issn = {1748-3395},
  doi = {10.1038/s41565-023-01583-0},
  copyright = {2024 The Author(s), under exclusive licence to Springer Nature Limited},
  langid = {english}
}

@article{yin_high-order_2023,
  title = {High-{{Order Exceptional Points}} in {{Pseudo-Hermitian Radio-Frequency Circuits}}},
  author = {Yin, Ke and Hao, Xianglin and Huang, Yuangen and Zou, Jianlong and Ma, Xikui and Dong, Tianyu},
  year = {2023},
  month = aug,
  journal = {Physical Review Applied},
  volume = {20},
  number = {2},
  pages = {L021003},
  publisher = {American Physical Society},
  doi = {10.1103/PhysRevApplied.20.L021003}
}

@article{zhang_topological_2025,
  title = {Topological {{Eigenvalue Braiding}} and {{Quantum State Transfer Near}} a {{Third-Order Exceptional Point}}},
  author = {Zhang, He and Liu, Tong and Xiang, Zhongcheng and Xu, Kai and Fan, Heng and Zheng, Dongning},
  year = {2025},
  month = may,
  journal = {PRX Quantum},
  volume = {6},
  number = {2},
  pages = {020328},
  publisher = {American Physical Society},
  doi = {10.1103/PRXQuantum.6.020328}
}

\end{document}